\documentclass[fleqn,usenatbib]{mnras}

\usepackage{newtxtext,newtxmath}

\usepackage[T1]{fontenc}

\DeclareRobustCommand{\VAN}[3]{#2}
\let\VANthebibliography\thebibliography
\def\thebibliography{\DeclareRobustCommand{\VAN}[3]{##3}\VANthebibliography}

\usepackage{graphicx}	
\usepackage{amsmath}	
\usepackage{gensymb}
\usepackage{symbols}
\usepackage{enumitem}
\usepackage{hyperref}
\usepackage{xspace}
\usepackage{orcidlink}
\usepackage{subcaption}

\title[HI discs as baryonic sub-grid physics probes]{HI discs of L$_{\ast}$ galaxies as probes of the baryonic physics of galaxy evolution}

\author[J. Gensior et al.]{
\parbox{\textwidth}{Jindra Gensior\orcidlink{0000-0001-6119-9883},$^{1}$\thanks{E-mail: jindra.gensior@uzh.ch}
Robert Feldmann\orcidlink{0000-0002-1109-1919},$^{1}$
Marta Reina-Campos\orcidlink{0000-0002-8556-4280},$^{2,3}$
Sebastian Trujillo-Gomez\orcidlink{0000-0003-2482-0049},$^{4,5}$
Lucio Mayer,$^{1}$
Benjamin W. Keller\orcidlink{0000-0002-9642-7193},$^{6}$
Andrew Wetzel\orcidlink{0000-0003-0603-8942},$^{7}$
J. M. Diederik Kruijssen\orcidlink{0000-0002-8804-0212},$^{8,9}$
Philip F. Hopkins\orcidlink{0000-0003-3729-1684}$^{10}$
and Jorge Moreno\orcidlink{0000-0002-3430-3232}$^{11,12}$
}
\\
$^{1}$Department of Astrophysics, University of Zurich, Winterthurerstrasse 190, 8057 Z{\"u}rich, Switzerland\\
$^{2}$Department of Physics \& Astronomy, McMaster University, 1280 Main Street West, Hamilton, L8S 4M1, Canada\\
$^{3}$Canadian Institute for Theoretical Astrophysics (CITA), University of Toronto, 60 St George St, Toronto, M5S 3H8, Canada\\
$^{4}$Astroinformatics Group, Heidelberg Institute for Theoretical Studies, Schloss-Wolfsbrunnenweg 35, 69118 Heidelberg, Germany\\
$^{5}$Astronomisches Rechen-Institut Zentrum für Astronomie der Universität Heidelberg, Mönchhofstraße 12-14, 69120 Heidelberg, Germany \\
$^{6}$Department of Physics and Materials Science, University of Memphis, 3720 Alumni Avenue, Memphis, TN 38152, USA\\
$^{7}$Department of Physics \& Astronomy, University of California, Davis, CA 95616, USA\\
$^{8}$Technical University of Munich, School of Engineering and Design, Department of Aerospace and Geodesy, Chair of Remote Sensing Technology, \\\hspace{2.2mm}Arcisstr. 21, 80333 Munich, Germany\\
$^{9}$Cosmic Origins Of Life (COOL) Research DAO, coolresearch.io \\
$^{10}$California Institute of Technology, TAPIR, Mailcode 350-17, Pasadena, CA 91125, USA\\
$^{11}$Department of Physics and Astronomy, Pomona College, Claremont, CA 91711, USA\\
$^{12}$Center for Computational Astrophysics, Flatiron Institute, 162 Fifth Avenue, New York, NY 10010, USA\\
}

\date{Accepted 2024 May 02. Received 2024 April 26; in original form 2023 September 30}

\pubyear{2023}

\begin{document}
\label{firstpage}
\pagerange{\pageref{firstpage}--\pageref{lastpage}}
\maketitle

\begin{abstract}
Understanding what shapes the cold gas component of galaxies, which both provides the fuel for star formation and is strongly affected by the subsequent stellar feedback, is a crucial step towards a better understanding of galaxy evolution. Here, we analyse the \hi properties of a sample of 46 Milky Way halo-mass galaxies, drawn from cosmological simulations (\emp~and \fc box). This set of simulations comprises galaxies evolved self-consistently across cosmic time with different baryonic sub-grid physics: three different star formation models [constant star formation efficiency (SFE) with different star formation eligibility criteria, and an environmentally-dependent, turbulence-based SFE] and two different feedback prescriptions, where only one sub-sample includes early stellar feedback. We use these simulations to assess the impact of different baryonic physics on the HI content of galaxies. We find that the galaxy-wide HI properties agree with each other and with observations. However, differences appear for small-scale properties. The thin \hi discs observed in the local Universe are only reproduced with a turbulence-dependent SFE and/or early stellar feedback. Furthermore, we find that the morphology of HI discs is particularly sensitive to the different physics models: galaxies simulated with a turbulence-based SFE have discs that are smoother and more rotationally symmetric, compared to those simulated with a constant SFE; galaxies simulated with early stellar feedback have more regular discs than supernova-feedback-only galaxies. We find that the rotational asymmetry of the \hi discs depends most strongly on the underlying physics model, making this a promising observable for understanding the physics responsible for shaping the interstellar medium of galaxies. 
\end{abstract}

\begin{keywords}
galaxies: evolution -- galaxies: formation -- galaxies: ISM -- galaxies: star formation  -- ISM: structure  
\end{keywords}



\section{Introduction}
The physics of star formation and stellar feedback present one of the largest uncertainties in our understanding of galaxy formation and evolution \citep[e.g.][]{Somerville2015,Naab2017,Crain2023}. An additional complication from a theoretical perspective is how to encode these small-scale processes in models for cosmological-scale simulations that cannot accurately resolve these scales. 

On galactic scales, the star formation rate (SFR) surface density is proportional to the (molecular) gas surface density \citep[e.g.][]{Kennicutt1998,Bigiel2008,delosReyes2019,Sun2023}. Gas is converted into stars with an efficiency of $\sim$1~per~cent \citep[e.g.][]{Leroy2008,Krumholz2012} per free-fall time ($\tff$, the time it would take for a gas cloud to collapse under its own self-gravity). However, there is growing evidence that the star formation efficiency (SFE) varies between galaxies (e.g.\ \citealt{Utomo2018}, \citealt{Chevance2020}, \citealt{Kim2022}, \citealt{Sun2023}), as well as within galaxies \citep[e.g.][]{Longmore2013,Kruijssen2014,Usero2015,Barnes2017,Querejeta2019}. This suggests that the SFE per free-fall time ($\eff$) is not constant (as is often assumed for the sub-grid modelling of star formation), but depends on properties of the local galactic environment on giant molecular cloud scales.

Analytic theory and high-resolution star formation simulations of molecular clouds in turbulent boxes suggest that $\eff$ depends on the turbulent properties of the gas. Specifically, on the virial parameter ($\avir$, the ratio of gravitational potential to turbulent kinetic energy of a cloud) and the turbulent Mach number ($\mn= \sigma/\cs$ the ratio of the turbulent velocity dispersion to the sound speed) of the gas \citep[e.g.][]{Krumholz2005,Padoan2011,Hennebelle2011,Federrath2012,Burkhart2018}. Such models can successfully explain the low star formation rate in the Milky Way \citep{Evans2022} and are able to reproduce the suppressed star formation rates observed \citep[e.g.][]{Davis2014} in molecular gas-hosting early-type galaxies \citep[e.g.][]{Gensior2020,Kretschmer2020,Gensior2021}, something that constant $\eff$ models struggle to accomplish.

The necessity of stellar feedback in regulating star formation and producing realistic galaxies resembling those observed in the (local) Universe has long been established \citep[e.g.][for reviews]{Somerville2015,Naab2017}. Stellar feedback is generally included via the effect of supernovae, and only some simulations include additional `early' stellar feedback processes (i.e.~all processes like winds, radiation pressure and photoelectric heating and photoionization by massive OB stars that act before the first supernova explodes). However, recent observations have highlighted the importance of the so-dubbed early stellar feedback processes\footnote{For brevity, we use this terminology throughout the remainder of the paper to refer to the wind, radiation pressure, photoelectric heating and photoionization feedback from massive OB stars.}  in destroying molecular clouds (e.g.\ \citealt{Chevance2022}, \citealt{Kim2022},\citealt{Chevance2023} and simulations including early stellar feedback reproduce these short cloud lifetimes, e.g.\ \citealt{Benincasa2020}, \citealt{Semenov2021}, \citealt{Keller2022}). However, incorporating parametrisations of these processes into simulations seems to primarily affect the density structure of the interstellar medium (ISM) and subsequent outflow mass loading, rather than the global star formation rate \citep[e.g.][]{Stinson2013,Smith2021,Keller2022}. 

How these more empirically-motivated star formation and stellar feedback models affect galaxy properties when taking into account their highly non-linear interplay through the evolution across cosmic time has been tested for individual objects in cosmological zoom-in simulations \citep[e.g.][]{Kretschmer2020,NunezCastineyra2021}. However, a thorough exploration of their effects in a statistical sample of objects has yet to be carried out. Combining the \emp~\citep{ReinaCampos2022} suite of cosmological zoom-in simulations and galaxies in the same halo mass range from the \fc box \citep{Feldmann2023} cosmological volume puts us in an ideal position to do so. Between them, our data-set consists of galaxies simulated with three different star formation models (constant $\eff$ with two different sets of criteria for making gas star-forming eligible, and a turbulence-based $\eff$) and two different stellar feedback models (only \fc box includes early stellar feedback). In this paper, we focus on analysing and comparing the \hi properties of these galaxies at $z=0$.

While molecular gas appears to be the predominant fuel for star formation \citep[e.g.][and references therein]{Tacconi2020}, \hi plays an important role as a gas reservoir for future episodes of star formation \citep[e.g.][]{Popping2014,Wang2020a,Saintonge2022}. Furthermore, the \hi discs of galaxies are strongly affected by the resultant stellar feedback: supernovae drive turbulence, enhancing the velocity dispersion of the \hi \citep[e.g.][]{Bacchini2020} and have the capacity to drive outflows and create massive holes in the \hi disc \citep[e.g.][]{Silich2001,Boomsma2008,Orr2022}. In addition to being intimately linked to both of the physical processes we are interested in, \hi is a well-studied ISM tracer in the nearby universe. Data on the properties of \hi discs are available through surveys such as the Westerbork HI survey of SPiral and irregular galaxies \citep[WHISP;][]{Swaters2002}, The \hi Nearby Galaxy Survey \citep[THINGS;][]{Walter2008}, the {\sc{bluedisk}} project \citep{Wang2013}, xGASS \citep{Catinella2018} and the pilot surveys of the Square Kilometer Array (SKA) precursors, such as the MeerKAT International GHz
Tiered Extragalactic Exploration \citep[MIGHTEE;][]{Jarvis2016} on MeerKAT and the Widefield ASKAP L-band Legacy All-sky Blind surveY \citep[WALLABY;][]{Koribalski2020} on the Australian SKA Pathfinder (ASKAP). Therefore, \hi (and the \hi disc properties in particular) promise to be an excellent observable against which to compare the simulated galaxies in our sample and to make predictions for the (precursors of the) SKA.  

The remainder of this paper is structured as follows. In Section~\ref{s:sims} we introduce our sample of simulated galaxies and the baryonic physics with which the simulations were run. We present and discuss the properties of the \hi discs of the galaxies in Section~\ref{s:HIprops}. In Section~\ref{s:HImorph}, we analyse the morphologies of the \hi discs, quantified with the non-parametric morphological indicators Gini, Smoothness and Asymmetry. Moving beyond a descriptive analysis, we utilise random forest regressions to infer the (galaxy) properties most relevant to predict the different HI disc morphologies in Section~\ref{s:rfr}. Finally, we conclude in Section~\ref{s:conclusion}.

\section{The simulations}\label{s:sims}

\subsection{EMP-\textit{Pathfinder}}\label{ss:EMP}
\emp~is a suite of cosmological zoom-in simulations of $\Ls$ galaxies introduced in \citet{ReinaCampos2022}, run with the moving-mesh code {\sc{arepo}} \citep{Springel2010,Weinberger2020} and the \emp~sub-grid physics implementation. Initial conditions match those of the MOdelling Star cluster population Assembly In Cosmological Simulations within EAGLE (E-MOSAICS; \citealt{Pfeffer2018,Kruijssen2019}) project and were drawn from the EAGLE Recal-L025N0752 DM-only periodic volume \citep{Schaye2015}. The E-MOSAICS initial conditions were selected solely for their halo mass, $11.85 < \log_{10}\left(M_{200}/{\rm M}_\odot\right) < 12.48$, to represent present-day Milky Way-mass galaxies. 

To match E-MOSAICS, \emp~has a baryonic mass resolution of $\sim2.2\times10^5~\Msun$, and $1\times10^6~\Msun$ for the highest-resolution dark matter particles, which populate the 600 kpc of the simulation volume surrounding the central galaxy. Gas softening is adaptive, with a minimum gravitational softening length of 56.3 c$\pc h^{-1}$, which is comoving for the entire run. The Plummer-equivalent gravitational softening for stars and dark matter is fixed in comoving units until $z=2$ (450 and 822 c$\pc h^{-1}$, respectively), and fixed to 175 and 320 pc, respectively at $z \leq 2$. 

The Grackle chemistry and cooling library\footnote{\href{https://grackle.readthedocs.io/}{https://grackle.readthedocs.io/}} \citep{Smith2017} with the 6-species chemistry network is used to model the thermal state of the interstellar medium. Specifically, the tabulated metal cooling, non-equilibrium chemistry for H, $\rm H^+$, He, $\rm He^{+}$, $\rm He^{++}$ and electrons, and photoelectric heating and photoionization from the \citet{Haardt2012} UV-Background, allow a self-consistent modelling of the multi-phase interstellar medium in the temperature range $10{-}10^9~\K$. 

An outstanding feature of \emp~relevant to this work is that it is a set of cosmological zoom-in simulations that evolved a suite of identical initial conditions with two different star formation sub-grid models, allowing us to study the effects that different sub-grid star-formation physics have when evolving galaxies self-consistently across cosmic time in a representative sample of Milky Way-halo mass galaxies rather than for individual objects. The fiducial sub-set of the sample (\empc), has $\eff = 20$~per~cent, while in \empv~the star formation efficiency per free-fall time depends on the turbulent state of the gas via the Mach number and the virial parameter. 
Specifically, the \citet{Federrath2012} multi-free-fall description of the \citet{Krumholz2005} model is used, following \citet{Kretschmer2020}. The SFE is given by:
\begin{equation}
    \centering
    \eff = \frac{1}{2}\exp\left(\frac{3\sigma_{s}^2}{8}\right)\left[1+\erf\left(\frac{\sigma_{s}^2-\scr}{\sqrt{2\sigma_{s}^2}}\right)\right],
\end{equation}\label{eg:SFEff}
where $\sigma_{\rm s} = \ln(1+0.49\mn^2)$ is the the width of the turbulent density probability distribution function and $\scr$ is the lognormal critical density for star formation
\begin{equation}
    \centering
    \scr = \ln \left[ \avir \left( 1 + \frac{2\mn^4}{1+\mn^2} \right) \right].
\end{equation}
The velocity dispersion (and thus Mach number) and virial parameter are calculated on the cloud scale using the overdensity method of \citet{Gensior2020}, which essentially performs an on-the-fly `cloud' identification for each gas cell.
Gas particles are eligible for star formation when their hydrogen number density exceeds 1 atom per cubic centimetre and their temperature is below $1.5\times10^4~\K$. This high temperature threshold is necessitated by initialising the simulation with gas particles that have primordial metallicities. The gas thus cannot cool below $\sim10^4~\K$ until the first stars have formed and subsequent supernovae and winds have enriched the remaining gas. However, once the the gas has been metal enriched, it can cool rapidly and stars form predominantly in gas with temperatures $\lesssim 200~\K$. Star formation is treated stochastically, and the main impact of the turbulence-based SFE is that star formation preferentially occurs at (much) higher gas densities, compared to \empc. This will be discussed in more detail in a dedicated \emp~star formation paper (Gensior et al. in preparation).
The stellar feedback channels included in \emp~are mass, metal, energy and momentum injection from supernovae of Type II and Type Ia, and winds from evolved (AGB) stars. Haloes were identified using a combination of the Friend-of-Friends (FoF; \citealt{Davis1985}) and {\sc{subfind}} algorithms \citep{Springel2001,Dolag2009}. 

\subsection{FIREbox}\label{ss:FB}
\fc box \citep{Feldmann2023} is a (22.1 Mpc)$^3$ cosmological volume simulation that is part of the Feedback In Realistic Environments (\fc\footnote{\href{https://fire.northwestern.edu/}{https://fire.northwestern.edu/}}) project. It was run with the meshless-finite-mass code {\sc{gizmo}}\footnote{\href{http://www.tapir.caltech.edu/~phopkins/Site/GIZMO.html}{http://www.tapir.caltech.edu/~phopkins/Site/GIZMO.html}} \citep{Hopkins2015} and the \fc-2 \citep{Hopkins2018} sub-grid physics implementation. We use the fiducial \fc box run (FB1024) for our analysis. This simulation has a baryonic mass resolution of $6.4\times10^{4}~\Msun$ and a dark matter mass resolution of $3.4\times10^5~\Msun$. The fiducial \fc box and its next lower resolution re-run (FB512) bracket the mass resolution of \emp, being a factor of 2 higher and lower respectively. We use the higher resolution \fc box, but note that our conclusions will not change when using the lower-resolution volume, due to the excellent convergence of the \hi properties of \fc box \citep[see e.g.\ online supplementary material in][]{Gensior2023a}. Gas particles have adaptive gravitational softening with a minimum Plummer-equivalent gravitational softening length of 1.5 pc (although the average softening length of star-forming gas is $\sim$20 pc), the Plummer-equivalent softening lengths of stars and dark matter particles are 20 pc and 80 pc respectively. All softening lengths are comoving at $z\geq 9$, and fixed in physical units for $z\leq 9$.   

The \fc-2 sub-grid physics model naturally leads to a multiphase ISM, including the cold phase, using the \citet{Hopkins2014} heating and cooling rates which are valid for temperatures ranging from 10 {--} $10^9~\K$ and accounting for photoelectric heating and photoionization from the \citet{FaucherGiguere2009} UV-Background. Star formation proceeds in gas above a density threshold of 300 atoms per cubic centimetre, that is self-gravitating ($\avir < 1$), Jeans unstable (Jeans mass lower than the gas particle mass), and molecular, with a star formation efficiency per free-fall time of 100~per~cent. While differing from \empv~in the details of the numerical implementation and the behaviour at higher Mach numbers, the \fc-2 star formation model is also motivated by turbulent star formation theory \citep[see e.g.][and discussion therein]{Hopkins2013,Hopkins2018}. \fc-2 includes the same stellar feedback channels as \emp, namely supernovae Type II and Ia and stellar winds from AGB stars. However, in addition \fc-2 also includes early stellar feedback from young massive stars, in the form of stellar winds, photoionization, photoelectric heating and radiation pressure from OB stars. 
Haloes were identified at $z=0$ using the AMIGA halo finder \citep{Gill2004,Knollmann2009}.

\subsection{The sample}\label{ss:sample}
Table~\ref{tab:sim_props} summarises the basic properties of the different simulation sub-sets. The initial conditions for the \emp~galaxies were selected to have halo masses in the range $11.85 \leq \log (\Mh/\Msun)\leq 12.3$ at $z=0$, comparable to that of the Milky Way \citep[e.g.][]{Bland-Hawthorn2016}. We thus apply the same halo mass cut to select the \fc box galaxies for this analysis. The full sample comprises 21 \empc, 14 \empv~and 26 \fc box central galaxies. However, some of these galaxies are visually classified as undergoing a major merger or interaction at $z=0$, hence we exclude them from the analysis. The reduced sample consists of 14 \empc, 12 \empv~and 20 \fc box galaxies. 

\subsubsection{Calculating $f_{\rm HI}$}\label{ss:calcfHI}

Although all simulations track `HI', \emp~via the Grackle non-equilibrium chemistry network and \fc box based on {\sc{CLOUDY}} tables that include the contributions from the UV-Background and local radiation from stars \citep[see][for details]{Hopkins2014,Hopkins2018}, this encompasses the entire neutral hydrogen phase, i.e.\ \hi and H$_2$. None of the simulations include non-equilibrium chemistry for H$_2$. Thus, we use two empirically-motivated models, one for each simulation, to estimate the molecular fraction of the gas and then subtract it from the total neutral gas fraction, to obtain the `true' \hi fraction of each gas particle.

We use the empirical \citet{Blitz2006} scaling relation between the gas pressure and H$_2$ to \hi ratio to determine $\MHI$ for the \emp~galaxies. Specifically, they found that
\begin{equation}
   R_{\rm mol} \equiv \frac{\SHt}{\SHI}  = \left(\frac{P}{P_0}\right)^\alpha,  
\end{equation}\label{eq:Rmol} where P is the mid-plane pressure of the gas, and $P_0$ and $\alpha$ are free parameters, calibrated from the observations of nearby galaxies. Following \citet{Marinacci2017}, we use the \citet{Leroy2008} values of $P_0 = 1.7\times10^4~\K~\ccm$ and $\alpha = 0.8$. The molecular fraction of neutral gas $f_{\rm H_2, neutral}$ is then given by $R_{\rm mol}/(R_{\rm mol}+1)$ and the \hi mass of each gas particle can be calculated as $M_{\rm HI,i} = f_{\rm neutral, i} \times \left(1-f_{\rm H_2, neutral, i}\right)\times M_{i}$ from the particle mass $M_{i}$, the neutral hydrogen fraction $f_{\rm neutral, i}$ and the pressure $P_{i}$.

While not including non-equilibrium chemistry, the \fc-2 model uses Equation 1 of \citet{Krumholz2011} to estimate $f_{\rm H_2, neutral}$ at run-time. This is an analytical model calibrated with high-resolution radiative transfer, non-equilibrium H$_2$ chemistry simulations of an idealised spherical cloud embedded in a Lyman-Werner background. In the \citet{Krumholz2011} model, $f_{\rm H_2, neutral}$ depends on the dust optical depth and the metallicity of the cloud. The dust optical depth scales with the surface density of the gas, which is calculated from the volume density and the scale height (obtained via a Sobolev approximation on the density). 

We use two different models to estimate $f_{\rm H_2, neutral}$ for the \emp~and \fc box galaxies on account of the differences in the simulation setup. While the pressure-based \citet{Blitz2006} estimate is more empirically motivated, \fc box was calibrated using the \citet{Krumholz2011} model. Using the pressure-based approach for \fc box leads to lower $f_{\rm H_2, neutral}$ and thus higher \hi masses and surface densities for the \fc box galaxies ($\SHI$ of 20--30 $\Msun~\pc^{-2}$ in the central regions of all galaxies cf.~$\approx 10~\Msun~\pc^{-2}$ with the \citet{Krumholz2011} model). Conversely, using the \citet{Krumholz2011} model leads to lower $f_{\rm H_2, neutral}$ and higher \hi fractions for the \emp~galaxies, in particular for \empv, compared to the pressure-based recipe (also yielding $\SHI$ of 20--30 $\Msun~\pc^{-2}$ in the central regions of half the galaxies in the sample). This likely results from the coarser spatial resolution in the \emp~simulation compared to \fc box (minimum gas gravitational softening of 83.6 pc at $z=0$ compared to 4.2 pc respectively, see Table~\ref{tab:sim_props}). The optical depth estimates obtained for the high-density, high-pressure regions, which are especially prominent in the centres of \empv~galaxies, are likely underestimates, thus leading to an underestimation of $f_{\rm H_2, neutral}$. Since observations suggest that $\sim 10~\Msun~\pc^{-2}$ \citep[e.g.][]{Blitz2006,Bigiel2008} is a threshold above which gas tends to be molecular, we selected the model that best reproduces this behaviour for the simulated galaxies as the fiducial way to calculate the \hi fraction of the galaxies. The qualitative results of this paper remain the same, independent of which model is used; however, there are some quantitative differences, which we discuss in Appendix~\ref{A:H2conv}. 

The resultant \hi distributions in \emp~and \fc box are in good agreement, despite the different chemistry/cooling treatment and ways to calculate $\fHI$. The distribution in temperature is bimodal, with the majority of \hi in the warm phase (T > 5000 K).

\begin{table*}
    \centering
    \caption{Properties of the simulation sub-samples.}
    \begin{tabular}{l|c|c|c|c|c|c|c|c|c|c|c}
        \hline
         Name & $N_{\rm gal}$ & $m_{\rm b}$ & $m_{\rm DM}$ & $\epsilon^{\rm min}_{\rm gas}$ & $d^{\rm med}_{\rm HI}$ & $\epsilon_{\rm stars}$ & $\epsilon_{\rm DM}$ & $n_{\rm th}$ & other SF criteria & \hspace{-1.3cm}$\eff$ & \hspace{-1.3cm}$e_{\rm SN}$ \\
         & & ($10^4~\Msun$) & ($10^5~\Msun$) & (pc) & (pc) & (pc) & (pc) & ($\ccm$) &  & \hspace{-1.3cm}& \hspace{-1.3cm}(ergs)\\
         \hline
         \emp~& 21 / 14 & $22.6$ & $14.4$ & 83.6 & 145 & 175 & 320 & 1 & $T< 1.5\times10^4~\K$ & \hspace{-1.3cm}0.2 & \hspace{-1.3cm}$3\times10^{51}$\\
         $\eff=20\%$ & & & & & & & & & \hspace{-1.3cm}& \hspace{-1.3cm}\\
         \emp~& 14 / 12 & 22.6 & $14.4$ & 83.6 & 120 & 175 & 320 & 1 & $T<1.5\times10^4~\K$ & \hspace{-1.3cm}$f(\avir, \mn)$ & \hspace{-1.3cm}$3\times10^{51}$ \\
         $\eff=f(\avir, \mn)$  & & & & & & & & & \hspace{-1.3cm}& \hspace{-1.3cm}\\
         \fc box & 26 / 20 & 6.26 & 3.35 & 4.2 & 76 & 12 & 80 & 300 & $\avir<1$,$f_{\rm H_2}$, $M_{\rm J} < m_{\rm b}$  & \hspace{-1.3cm}1  &  \hspace{-1.3cm}$10^{51}$ \\
         \hline
\multicolumn{11}{p{\linewidth}}{\footnotesize{\textit{Notes:} For each baryonic physics sub-sample, column 1 lists the name of the sub-sample, column 2 lists the total / non-interacting number of galaxies in the sample, column 3 lists the baryonic mass resolution, column 4 the dark matter mass resolution, columns 5 lists the minimum gravitational softening for gas, column 6 lists the HI mass-weighted median inter-particle spacing in the central regions of the galaxies, columns 7 and 8 list the Plummer-equivalent gravitational softening for stars and dark matter (all at $z=0$), column 9 lists the density threshold for star formation, column 10 lists other star formation criteria, column 11 lists the star formation efficiency per free-fall time and column 12 lists the energy injected per supernova.}} \\
    \end{tabular}
    \label{tab:sim_props}
\end{table*}

\subsubsection{The HI main sequence}\label{ss:HIMS}

\begin{figure}
    \centering
    \includegraphics[width=\linewidth]{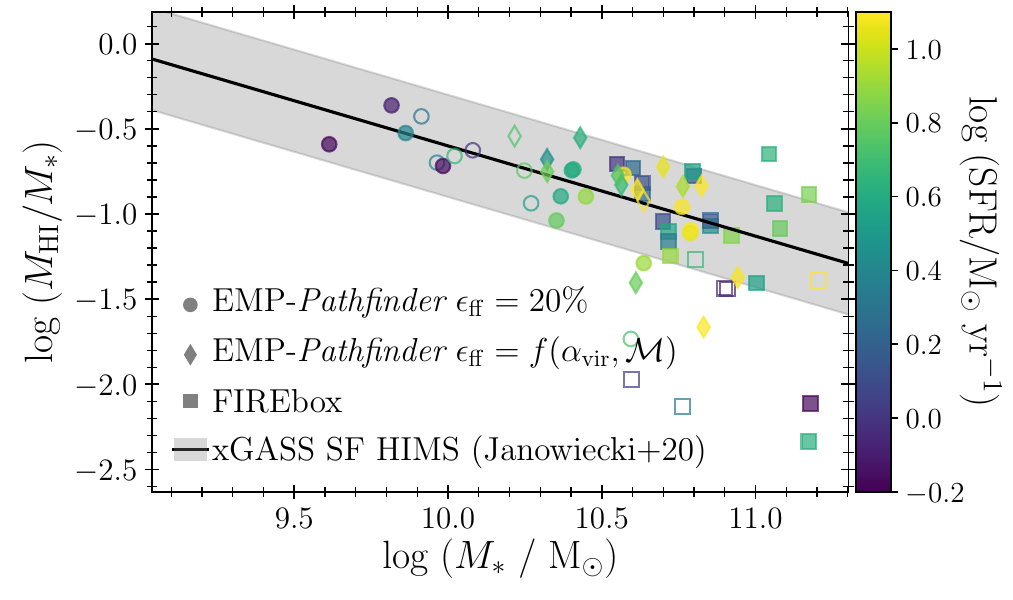}
    \caption{HI-to-stellar mass fraction as a function of stellar mass of the simulated galaxies in our sample. The data are colour-coded by their star formation rate, with different symbols denoting the different sub-grid physics samples. Empty symbols denote the galaxies excluded from further analysis, due to interactions. The data are overplotted on the \hi main sequence relation of the star-forming xGASS \citep{Catinella2018} galaxies from \citet{Janowiecki2020} shown as a solid black line, and the 0.3 dex uncertainty in grey shading.}
    \label{fig:HIMS}
\end{figure}

Figure~\ref{fig:HIMS} shows the HI-to-stellar mass ratio of the galaxies as a function of their stellar mass, colour-coded by their star formation rate. Filled symbols denote the galaxies considered for the analysis presented in the remainder of this paper, while the open symbols show the galaxies that are excluded due to interactions. The simulation data are overplotted on the xGASS \citep{Catinella2018} \hi `main sequence' of star-forming galaxies \citep{Janowiecki2020} shown as a black line with the 0.3 dex scatter around this \hi main sequence relation indicated through grey shading. Despite their similar halo masses at $z=0$, the galaxies evolved with the different sub-grid physics models are clustered in different parts of in the \hi-to-stellar mass fraction - stellar mass plane, highlighting the impact of subtle differences in baryonic physics on galaxy evolution across cosmic time. While there is overlap between all simulation sub-sets around the stellar mass of the Milky Way $\log(\Mstar/\Msun)\sim10.7$ \citep[e.g.][]{Cautun2020}, the \empc~galaxies tend to be undermassive (10 galaxies with $\log(\Mstar/\Msun)\leq10.3$) and the \fc box galaxies extend to higher stellar masses (8 galaxies with $\log(\Mstar/\Msun)\geq11$). The \hi-to-stellar mass fractions of all simulated galaxy sub-sets are in good agreement with the observations, scattering around the \hi main sequence mostly within the $\pm$0.3 dex scatter. 

\section{HI disc properties}\label{s:HIprops}
In this section, we examine the properties of the \hi discs in more detail, beginning with properties that have been observationally well-established. To do so, we rotate all galaxies such that the \hi disc is face-on, i.e.~in the $x${--}$y$ plane and the angular momentum vector of the gas, calculated based on all gas within the central 5 kpc, is perpendicular to it. Specifically, we examine the \hi mass-size relation (Section~\ref{ss:HISM}), radial \hi surface density and scale height profiles (Sections~\ref{ss:HISD} and Sections~\ref{ss:HIsh}, respectively).

\subsection{HI mass-size relation}\label{ss:HISM}
The \hi mass-size relation relates the mass of \hi enclosed within the \hi scale radius (R$_{\rm HI}$, the radius of the \hi disc where $\Sigma_{\rm \hi} = 1~\Msun\pc^{-2}$) to the diameter of the \hi disc ($D_{\rm \hi}$ = 2 R$_{\rm HI}$). It is a very tight relation that has a slope of $\sim$0.5, and has been well established empirically \citep[e.g.][]{Broeils1997, Swaters2002, Begum2008, Lelli2016, Wang2016, Stevens2019, Rajohnson2022}. It can serve as a test for the (baryonic physics) of cosmological simulations: due to the robustness of the \hi mass-size relation, simulated galaxies should lie on the relation, unless the star formation and/or stellar feedback physics result in very disturbed \hi disc morphologies containing enormous `holes' (\citealt{Stevens2019}, see also the EAGLE \hi mass-size relation and the discussion regarding unphysically large \hi holes in \citealt{Bahe2016}). 

\begin{figure}
    \centering
    \includegraphics[width=\linewidth]{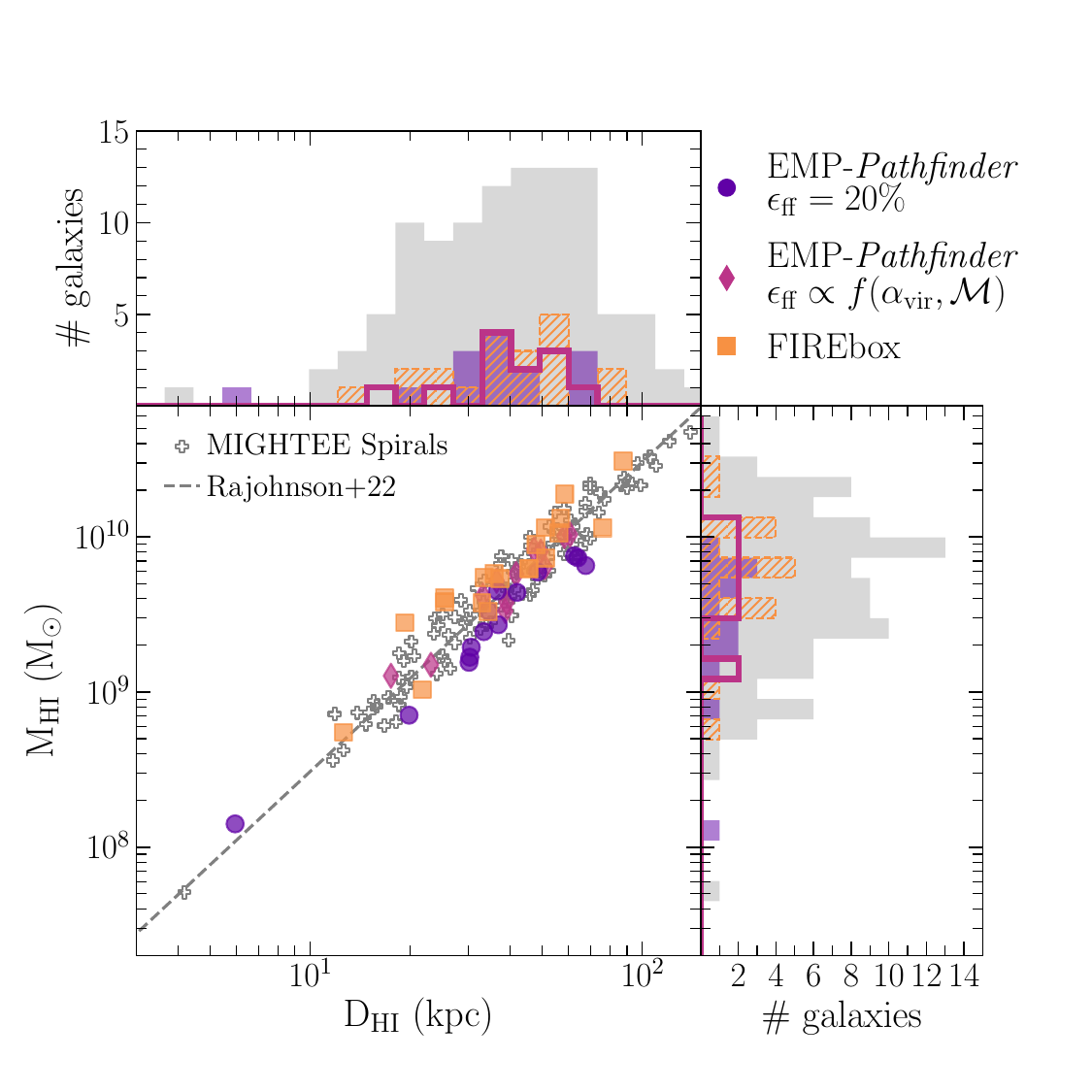}
    \caption{\hi mass-size relation for the galaxies in our sample. Purple circles denote the \empc~galaxies, magenta diamonds denote the \empv~galaxies, and orange squares denote the \fc box galaxies. Histograms show the marginal distributions for each indicator and galaxy sub-set. The grey crosses show the observational data from MIGHTEE, the grey dashed line shows their best-fit HI mass-size relation \citep{Rajohnson2022}. All simulated galaxies scatter tightly around the observed HI mass-size relation.}
    \label{fig:HIMD}
\end{figure}

The main panel of Figure~\ref{fig:HIMD} shows the \hi mass-size relation for our sample of simulated galaxies, with histograms showing the marginal distributions of $\MHI$ and $D_{\rm HI}$. Grey crosses show the data for MIGHTEE spirals, and the dashed line shows the most recent empirical fit for the \hi mass-size relation to this data \citep{Rajohnson2022}. To first order, all galaxy sub-sets lie on the \hi mass-size relation, with a scatter comparable to that of the observations. The \empv~galaxies are in excellent agreement with the MIGHTEE fit. The \empc~galaxies tend to have slightly under-massive \hi discs for their sizes, the opposite holds for \fc box galaxies, which tend to be a little over-massive at a fixed $D_{\rm HI}$, compared to the MIGHTEE fit. 

\subsection{HI surface density profiles}\label{ss:HISD}
Next, we turn to the radial \hi surface density profiles, which we compute in rolling bins of width 1 kpc. In the outskirts of the \hi disc ($R/R_{\rm HI} \gtrsim 0.8$), the $\SHI$ profiles of observed late-type and low-mass galaxies decline exponentially \citep[e.g.][]{Swaters2002, Obreschkow2009, Bigiel2012, Wang2014}. The shape and normalisation of the inner profile tend to vary between galaxies, observational samples and morphological types, with central median $\SHI$ ranging from 4 to 8 $\Msun\pc^{-2}$ for dwarfs and late-type galaxies with stellar masses to $10^{11}~\Msun$ \citep[][in particular their Figure 2 for a compilation of median profiles from observational surveys, and references therein]{Wang2016}. 

\begin{figure}
    \centering
    \includegraphics[width=\linewidth]{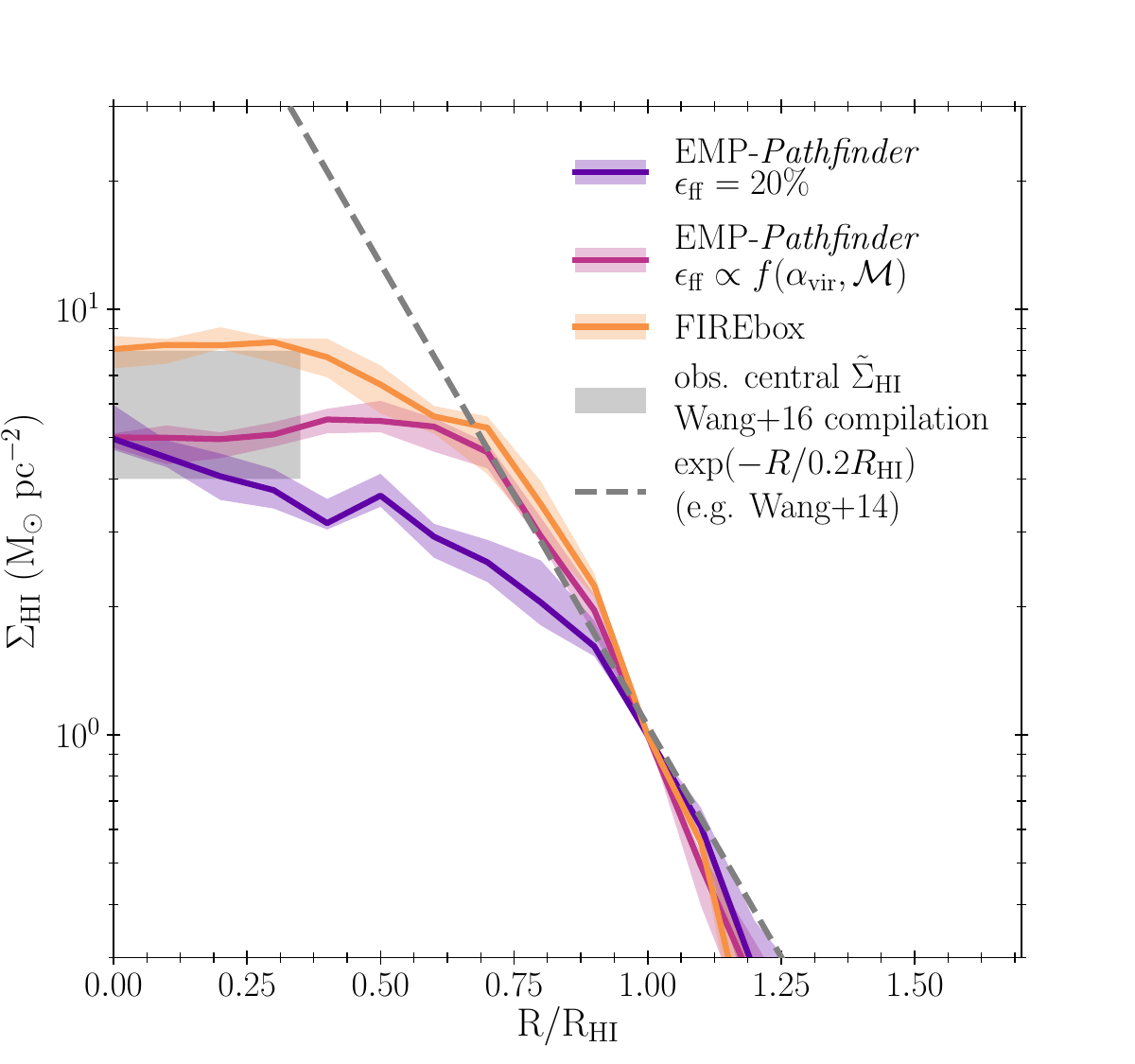}
    \caption{\hi gas surface density profiles as a function of galactocentric radius scaled by the \hi scale radius $R_{\rm HI}$ (defined as radius where $\Sigma_{\rm \hi} = 1~\Msun\pc^{-2}$). Coloured lines and shaded regions denote the median and the error on the median, determined via bootstrapping, of each different baryonic physics sub-sample, respectively, with the \citet{Wang2014} exponential fit to the outer regions of observed HI surface density profiles overplotted as a grey-dashed line. For comparison, the grey shaded box indicates the central range of the median $\SHI$ from 8 different observational samples, compiled by \citet{Wang2016}. While all simulated galaxy sub-sets approximately follow the exponential profile at $R \gtrsim 0.8R_{\rm HI}$, the central \hi mass surface density differs in both shape and normalisation.}
    \label{fig:HISD}
\end{figure}

Figure~\ref{fig:HISD} shows the median radial \hi surface density profiles of our galaxy sub-sets, the shaded regions indicating the error on the median determined via bootstrapping, with the \citet{Wang2014} exponential trend of the outer $\SHI$ profile overplotted as a grey-dashed line and the range of median central \hi surface densities of the observational compilation in \citet{Wang2016} indicated by a grey box. All median profiles follow the exponential profile in the outskirts of the disc (\empv~and \fc box for  $R \geq 0.8R_{\rm HI}$, \empc~only for $R\geq R_{\rm HI}$), however, the inner profiles differ both in shape and normalisation. \empc~has a centrally peaked median \hi profile (5 $\Msun\pc^{-2}$), which declines by  $\sim4~\Msun\pc^{-2}$ before dropping off exponentially. By contrast, the \empv~profile is approximately constant at $\sim5~\Msun\pc^{-2}$ as a function of radius, before declining exponentially at $R> 0.8 R_{\rm HI}$. The median \fc box $\SHI$ profile is approximately constant at 8 $\Msun\pc^{-2}$ to $R = 0.3 R_{\rm HI}$, before gradually declining to 5$\Msun\pc^{-2}$ at $R = 0.8 R_{\rm HI}$ and declining exponentially at larger radii. The central profiles of all simulated galaxy sub-sets fall within the profile shapes and magnitudes found within galaxies in the local Universe.

\subsection{HI scale heights}\label{ss:HIsh}
\begin{figure}
    \centering
    \includegraphics[width=\linewidth]{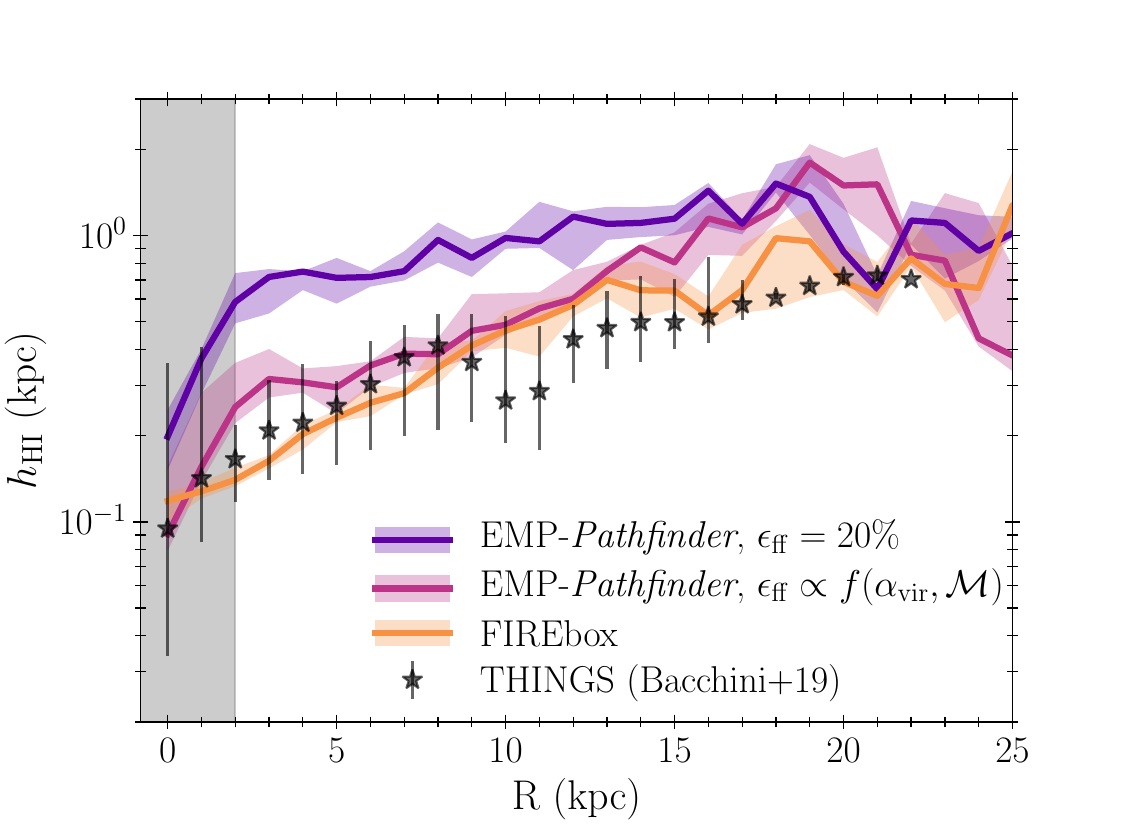}
    \caption{Radial profiles of the \hi disc scale heights of the simulated galaxies, measured by fitting a Gaussian to the vertical volume density distribution in kpc bins. Coloured lines and shaded regions denote the median and the error on the median, determined via bootstrapping, of each different baryonic physics subsample, respectively, with the THINGS scale heights \citep{Bacchini2019a} overplotted as black stars. The grey shaded region indicates the central region where scale heights are marginally resolved. The \empc~\hi discs are substantially thicker than those observed in the nearby universe within the inner $\sim$10 kpc.}
    \label{fig:hHI}
\end{figure}

Following \citet{Gensior2023a}, we compute the \hi scale height of the simulated galaxies in radial annuli of 1 kpc width, in the 4 kpc surrounding the galactic mid-plane. The vertical, volumetric gas density distribution in each annulus is fit with a Gaussian profile, $\rho(z) \propto \exp(-z^2/(2\hhi^2))$, which depends on the \hi scale height, $\hhi$. Figure~\ref{fig:hHI} shows the median radial \hi scale height profiles for the different sub-grid physics sub-sets, with the \citet{Bacchini2019a} scale heights for 12 THINGS galaxies, computed by iteratively fitting the vertical volume density profile (estimated from the total gravitational potential) with a Gaussian, overplotted. 

All simulated \hi discs increase in thickness from a median of $\sim$ 100 pc (\empv~and \fc box) {--} 200 pc (\empc) in the centre to $\sim$ 1 kpc in the outskirts of the disc, i.e.\ exhibit flaring. In contrast to the other two simulated galaxy sub-sets, the median scale height of the \empc~galaxies increases steeply from $\sim$ 200 pc to $\sim$ 700 pc within the central 3 kpc, before rising more shallowly to 1 kpc at larger radii. The median $\hhi$ of the \empv~galaxies also increases moderately from $\sim$ 100 pc to $\sim$ 300 pc within the central 3 kpc before increasing more gradually, compared to the \fc box galaxies' $\hhi$, which increases gradually throughout the disc. The median \hi scale heights of both \empv~and \fc box galaxies are in good agreement with the THINGS observations within the inner 10 {--} 15 kpc, with only the \fc box galaxies matching the observations with scale heights of several hundred pc in the outskirts of the \hi discs. The discs of the \empc~galaxies are consistently thicker than those of the other sub-sets of simulated galaxies and observations (compared to both THINGS \citep{Bacchini2019a, Patra2020}, but also the {\sc{bluedisk}} galaxies \citep{Randriamampandry2021}, which have similar scale heights). 

High resolution has been important for reproducing thin discs in simulations (e.g. \citealt{Guedes2011}, \citealt{Pillepich2019}, see also discussion in \citealt{Gensior2023a}). \fc box has a minimum gravitational softening of 4.2 pc for the gas and a Plummer-equivalent softening of 12 pc for the stars, which suggests that gravity around the mid-plane is well resolved in both the gaseous and stellar component. The minimum gravitational softening length for gas in \emp~is 83.6 pc (and the approximate gas cell radius will be a factor of $\sim$3 smaller per definition), while the Plummer-equivalent softening of the stars is 175 pc. 
Thus, the stellar density near the mid-plane might be underestimated. However, the median \hi scale height of \empc~galaxies enters the spatial regime where stellar gravity is well resolved ($\hhi > \epsilon_{\ast}$ ) for $R>1.5~\kpc$, while significant differences to the (much) lower median \empv~$\hhi$, which is broadly consistent with both \fc box and the observations, persist out to $R=15~\kpc$. Since the densest gas is likely pre-dominantly molecular, the HI mass-weighted median inter-particle spacing, $d_{\rm i} = (m_{\rm i} / \rho_{\rm i})^{1/3}$, calculated from the mass and density of each particle, can give a better idea of the characteristic size of HI-dominated particles. The inter-particle spacing in the central region is listed in column 6 of Table~\ref{tab:sim_props} and is 76 pc for \fc box, 120 pc for \empv~ and 145 pc for \empc, respectively. This indicates that the scale heights in the central 1--2 kpc are marginally resolved, shown as the grey-shaded region in Figure~\ref{fig:hHI}, while $\hhi$ will be resolved by several cells at larger galacocentric radii. Therefore, this implies that the underlying physics is the driver of the \hi scale height trends, and that the (\emp) resolution adequately resolves the scale heights for $R\gtrsim2~\kpc$.

\begin{figure}
    \centering
    \includegraphics[width=\linewidth]{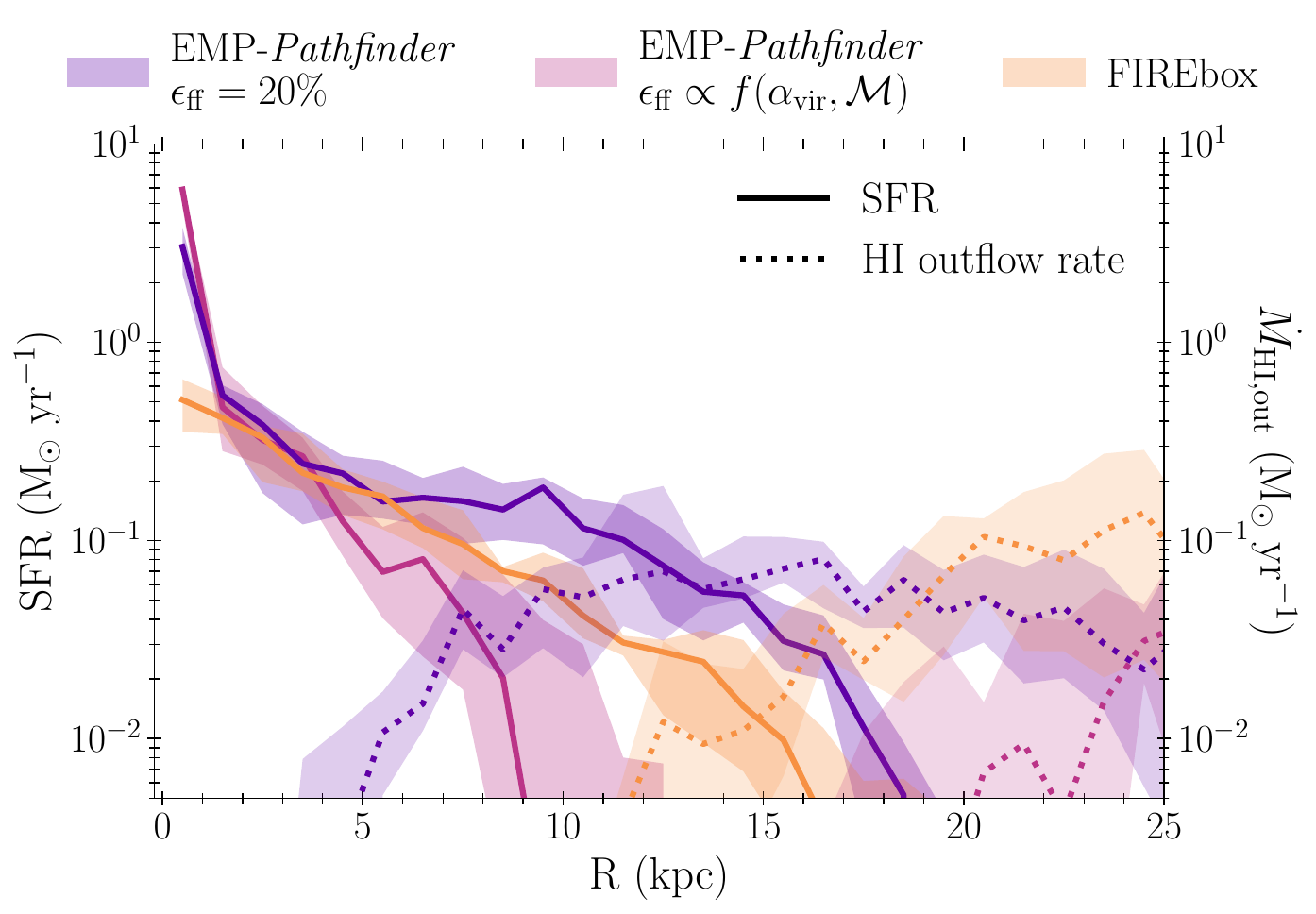}
    \caption{Radial profiles of the median star formation rates (averaged over a period of 100 Myr, solid lines) and \hi outflow rates (HI mass moving away from the galactic midplane in a 500 pc wide slab 5 kpc above and below the midplane, dotted lines) for the different simulated galaxy sub-sets. Coloured lines and shaded regions denote the median and the error on the median of each property, determined via bootstrapping, respectively. }
    \label{fig:outflows}
\end{figure}

Analysing (different ways to calculate the) \hi scale heights in \fc box galaxies, \citet{Gensior2023a} concluded that self-consistently modelling a multi-phase ISM that includes a cold phase could be a reason that the \fc box galaxies have thin \hi discs, in good agreement with observations. Comparing the observational results (and \fc box) to those of the \emp~galaxies, in particular to the \empc~galaxies which have discs that are consistently too thick, highlights that including a cold ISM is not a sufficient criterion for producing the thin \hi discs. Figure~\ref{fig:hHI} suggests that more empirically motivated sub-grid physics are required, either in the form of an environmentally-dependent SFE or early stellar feedback (which for the EMP family of simulations will be presented by Keller et al.~in prep.; Kruijssen et al.~in prep.). Both result in changes in location and strength of the stellar feedback compared to the \empc~case, reflected in the differences in the global galaxy, gas and star formation properties after an evolution across cosmic time.

We show the median radial profiles of the SFR, averaged over the past 100 Myr, and the \hi outflow rates, calculated as the mass moving away from the galaxy in a 500 pc wide slice located 5 kpc above and below the midplane, in Figure~\ref{fig:outflows}. The median SFR is much more centrally peaked in the \emp~galaxies compared to \fc box, which through the resultant stellar feedback likely leads to the steeper increase in $\hhi$ of \emp~galaxies within the inner 3 kpc. Additionally, the \empc~galaxies experience much stronger outflows in the central 15 kpc, compared to \empv~and \fc box galaxies, suggesting that the \empc~\hi discs are puffed up by feedback. This is in agreement with the results \citet{BenitezLlambay2018}, who found that the gas discs in EAGLE were too thick due to the strong stellar feedback. 
 
\section{HI disc morphologies}\label{s:HImorph}
We quantify the structure of the \hi discs using the non-parametric morphological indicators. 
These indicators are most commonly used to classify galaxy stellar morphologies in the optical \citep[e.g.][]{Abraham1994,Abraham1996,Conselice2003, Lotz2004, RodriguezGomez2019}. However, they have also been used to study the morphology of HI discs, particularly trying to identify mergers and interactions \citep[e.g.][]{Holwerda2011a,Holwerda2011b}, ram pressure stripping \citep[e.g.][]{Holwerda2023} and to assess how well property-matched IllustrisTNG50 galaxies resemble WHISP galaxies \citep{Gebek2023}. Furthermore, \citet{Davis2022} used the Gini, Asymmetry and Smoothness indicators to quantify the molecular gas morphologies in the central 3 kpc of a sample of late-and early-type galaxies from the mm-Wave Interferometric Survey of Dark Object Masses (WISDOM) project and Physics at High Angular Resolution in Nearby GalaxieS (PHANGS; \citealt{Leroy2021a}) survey and correlate the central ISM morphologies with galaxy properties. 

\subsection{Method}\label{ss:morphm}
Here we focus on the Asymmetry, Smoothness and Gini indicators to quantify the structure of the entire \hi disc of the galaxies. We utilise the ray-tracing capability of {\sc{arepo}} to generate \hi surface density projections for every simulated galaxy in the sample, including those from \fc box. Since the majority of \hi is in the warm phase, surface density maps should provide a reasonable estimate of the HI emission, but a full forward modelling with radiative transfer would be desirable for an one-to-one comparison with observations. The projections are generated face-on, in a box 60 kpc a side, centred on the centre of the galaxy. This is defined as the position of the particle with the minimum gravitational potential energy for \emp~haloes identified with {\sc{subfind}} and corresponds to the position where the total matter density is maximised for \fc box haloes identified with AMIGA. The \hi surface density maps have an intrinsic resolution of 20 pc per pixel. We then apply a Gaussian smoothing kernel with full-width half-maximum corresponding to 80 pc, comparable to the minimum gravitational softening length of the \emp~galaxies. Prior to computing the nonparametric morphological indicators, we apply a surface density cut corresponding to a HI column density of $7\times10^{19}~\rm cm^{-2}$, comparable in sensitivity to e.g.\ the {\sc{bluedisk}} \citep{Wang2013} or THINGS \citep{Walter2008} data and well within the sensitivity of SKA-precursor surveys MHONGOOSE \citep{deBlok2016} and MIGHTEE \citep{Maddox2021}. 

As a test for these choices, we also explore how a more conservative cut affects the results by recomputing the statistics for a cut at $3\times10^{20}~\rm cm^{-2}$, and at a coarser resolution of 500 pc. Furthermore, we also create maps at inclinations of 10, 30, 50 and 70$^{\circ}$ and measure Gini, Smoothness and Asymmetry. Notably, the qualitative trends discussed in Section~\ref{ss:morphr} are not strongly affected by these changes, as especially the Asymmetry varies little with inclination. We refer the reader to Appendix~\ref{A:GAS+} for a more extensive discussion. 

We follow \citet{Davis2022} and calculate the Asymmetry, $A$, as:

\begin{equation}
    \centering
    A \equiv \frac{\sum_{i,j} \left| I_{ij} - I^{180}_{ij}\right|}{\sum_{i,j} \left|I_{ij}\right|},
\end{equation}\label{eq:A}where $I_{ij}$ is the surface density of the pixel in position $ij$, and $I^{180}_{ij}$ the surface density of the pixel in the same position after the map has been rotated by 180 degrees, and the total Asymmetry\footnote{Using different definitions of $A$, such as replacing the denominator of Equation~\ref{eq:A} by $\sum_{i,j}  | I_{ij} + I^{180}_{ij} |$ \citep[e.g.][]{Lelli2014}, or squaring the difference instead of using absolutes \citep[e.g.][]{Deg2023}, affects the value of $A$, but has no impact on the results presented in this paper.} is obtained by summing over all pixel positions $i,j$ in the map. A lower value of $A$ corresponds to a more rotationally symmetric gas distribution, where $A=0$ indicates perfect rotational symmetry and $A=2$ complete asymmetry with all flux concentrated in one half of the map.

The Smoothness, $S$, is defined as:
\begin{equation}
    \centering
    S \equiv \frac{\sum_{i,j} \left| I_{ij} - I^{S}_{ij}\right|}{\sum_{i,j} I_{ij}},
\end{equation}\label{eq:S}where $I^{S}_{ij}$ is the surface density of the pixel $ij$ after it has been smoothed with a boxcar filter of width $\sigma$. In optical studies, the width of the boxcar filter is commonly set to 0.25 Petrosian radii and thus varies for each galaxy (e.g.\ \citealt{RodriguezGomez2019}), while \citet{Davis2022} use the same smoothing filter width for all galaxies. Following \citet{Davis2022}, we also use a single $\sigma$, i.e.\ fix the width of the boxcar filter for all galaxies. Specifically, we choose $\sigma = 4~\kpc$, which is still significantly larger than the 500 pc resolution of the coarser maps and comparable to a quarter of the Petrosian radius expected for discs this size. Furthermore, we have verified that varying the boxcar filter width only affects the absolute value of $S$ measured (larger smoothing kernel widths leading to higher values of $S$), but not the trends seen between the different physics sub-sets, in agreement with \citet{Davis2022} findings. Similar to the Asymmetry coefficient, a lower value of $S$ corresponds to a smoother gas distribution.

The Gini coefficient, $G$, is calculated as:
\begin{equation}
    \centering
    G \equiv \frac{1}{\bar{X} n (n-1)} \sum^{n}_{i=1} (2i - n - 1)X_i,
\end{equation}where $\bar{X}$ is the mean surface density measured over all pixels $i = 1, 2,...,n$ and $X_i$ the surface density of each individual pixel. The Gini coefficient measures how homogeneously the flux is distributed across the map, where a completely homogeneous brightness distribution yields $G=0$ and $G=1$ corresponds to all flux being concentrated in a single pixel. 

\subsection{Results} \label{ss:morphr}
\begin{figure*}
    \centering
    \includegraphics[width=\linewidth]{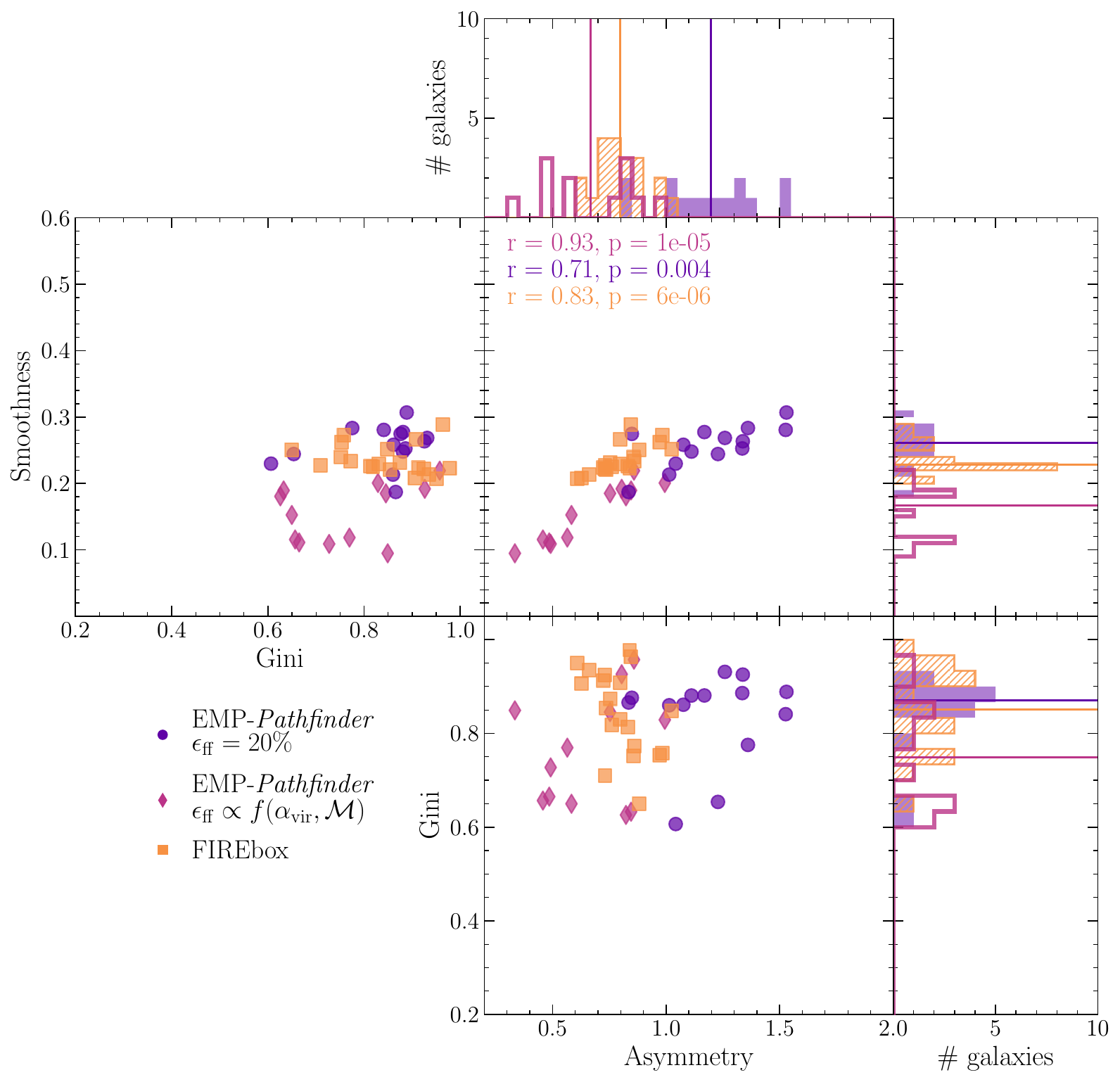}
    \caption{\hi morphology as classified by the non-parametric morphological indicators Smoothness, Gini and Asymmetry (plotted against each other in the three main panels) measured from face-on \hi surface density projections of the galaxies with a resolution of 80 pc and column density threshold of $7\times10^{19}~\rm{cm}^{-2}$. Purple circles denote the \empc~galaxies, while magenta diamonds denote the \empv~galaxies, and orange squares denote the \fc box galaxies. Histograms show the marginal distributions for each indicator and galaxy sub-set, the median values are indicated by coloured lines. A statistically significant correlation between Smoothness and Asymmetry is present for each simulated galaxy sub-sets, the Spearman rank correlation coefficient and $p$-value are indicated in the top left corner of the panel in the respective colour of each sub-set. No statistically significant correlations are present between Gini and Smoothness or Asymmetry. The galaxies simulated with different sub-grid physics occupy distinctly different parts of the Asymmetry-Smoothness parameter space.}
    \label{fig:HImorph}
\end{figure*}

The main panels of Figure~\ref{fig:HImorph} show the Smoothness, Gini and Asymmetry coefficients for the different simulated galaxy sub-sets, plotted against each other, while their marginal distributions are shown as histograms on the side. There is a strong correlation between Smoothness and Asymmetry for all individual sub-sets of galaxies simulated with different sub-grid physics and for the entire sample. No statistically significant correlation is present between Gini and Smoothness or Gini and Asymmetry for either the different simulated galaxy sub-sets or the entire sample. 

Remarkably, the different physics sub-sets occupy different parts of the parameter space in Smoothness and Asymmetry, demonstrating that the non-parametric HI morphology is a sensitive test of sub-grid physics in simulations. The \empv~galaxies are the smoothest and most symmetric (6 galaxies with $S<0.15$ and $A<0.6$), whilst the \empc~galaxies are the least smooth and most asymmetric (12 galaxies with $A>1$). \fc box galaxies occupy the intermediate parameter space between the \emp~galaxies, although there is significant overlap in the Smoothness of \fc box and \empc~galaxies ($0.2\leq S \leq 0.3$). 

All galaxies have Gini $\geq$ 0.6, with a large scatter. The high Gini values are likely a consequence of the \hi surface density distribution in these objects. Especially in the \empv~and \fc box case, the surface density profiles are roughly constant until they decline exponentially (see Figure~\ref{fig:HISD}), but even for \empc~there is a sharp contrast between the higher density regions in the \hi disc as the density rapidly drops below the density threshold. As Gini measures the inequality in the density distribution, it is therefore not surprising that it is more similar between the different physics sub-sets and is less sensitive to the disc morphology than Smoothness and Asymmetry.

\begin{figure*}
    \centering
    \includegraphics[width=\linewidth]{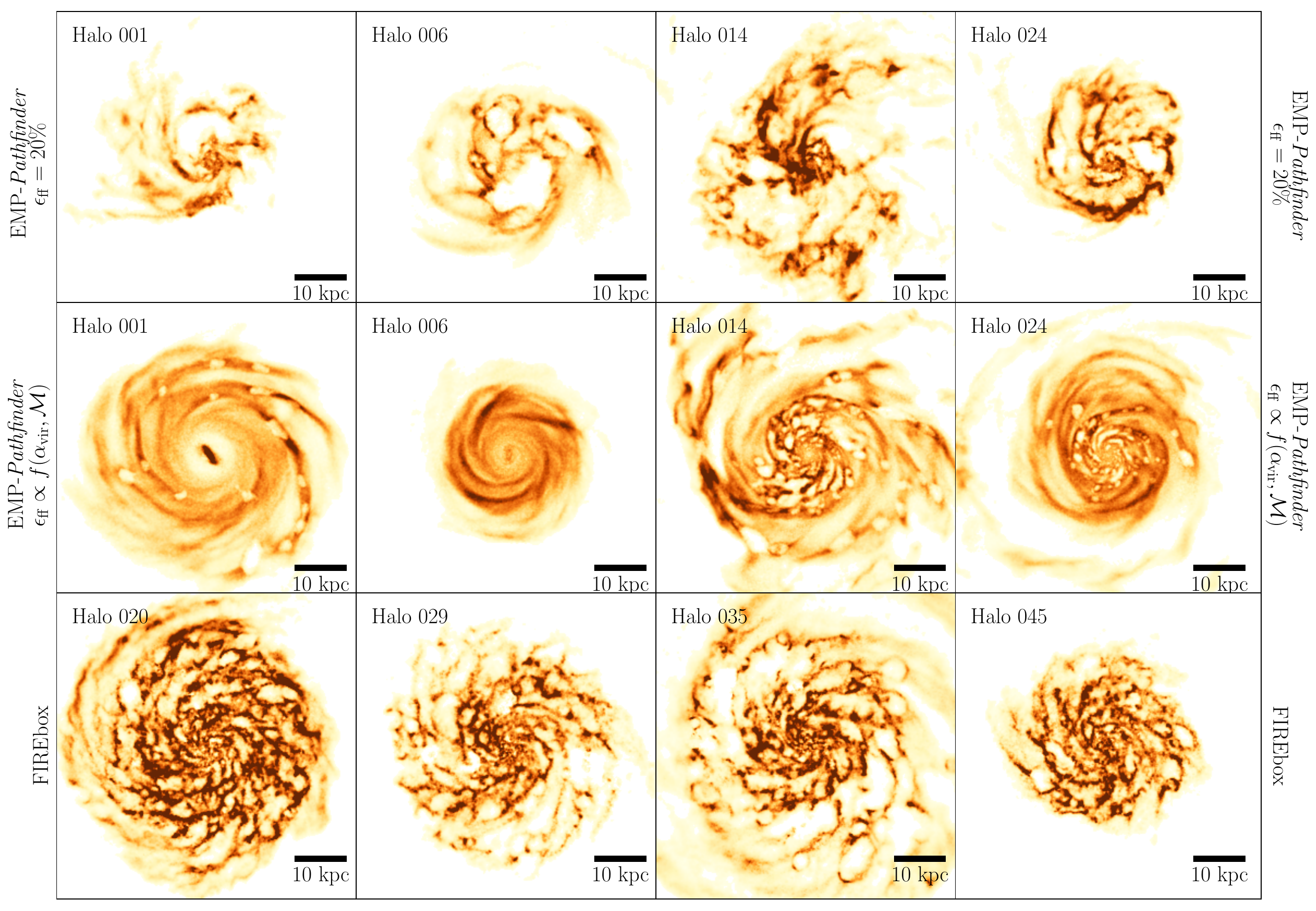}
    \caption{Face-on projection of the \hi surface density, 60 kpc a side, for four example galaxies of each baryonic physics galaxy sub-set. From top to bottom, the maps show galaxies evolved with the \empc, \empv~(where the same galaxies have been chosen to highlight the differences) and \fc box physics. These are the fiducial projections with a resolution of 80 pc and a density threshold of $7\times10^{19}$ cm$^{-2}$. The colourmap linearly scales with the gas surface density from light to dark brown at 10 $\Msun~\pc^{-2}$. The white background colour indicates the regions where the \hi surface density is below the threshold.}
    
    \label{fig:SDex}
\end{figure*}

None the less, Figure~\ref{fig:HImorph} (and Figures~\ref{fig:HImorph_hc} {--} ~\ref{fig:HImorph_inc}) highlight the impact of the different baryonic sub-grid physics on the distribution of \hi within the gas discs. We show some example surface density projections in Figure~\ref{fig:SDex}. The \empc~galaxies are characterised by large `holes'\footnote{Rather than being completely devoid of \hi the gas is at densities lower than our respective column density threshold.} within the \hi disc. These large holes, combined with asymmetrically distributed bright clumps of \hi result in the large Asymmetry and Smoothness values for the \empc~galaxies. \fc box galaxies are the most similar between them: all \hi distributions look similar to a flocculent spiral, with small holes as the result of stellar feedback. The \empv~sample is split between galaxies with similar morphologies to those of \fc box and the very smooth, fairly uniform density and hole-free galaxies that occupy the bottom left corner of the Smoothness-Asymmetry parameter space. 
Both \fc box and \empc~allow star formation anywhere within the disc (wherever the density threshold is met, which for \fc~ usually means that the other eligibility criteria are also met, see \citealt{Hopkins2018}). The only stellar feedback channel in \emp~galaxies is the SN feedback, leading to huge bubbles in the \empc~galaxies. The early stellar feedback in \fc box likely pre-processes the ISM in the natal environment of the stars, making it easier for the SN feedback to escape vertically rather than expanding into the plane of the galaxy, thus leading to smaller holes (which is qualitatively similar to the behaviour of \emp~combined with early stellar feedback in isolated galaxies, see Figure 8 in \citealt{Keller2022}). Contrary to that, the turbulence-based star formation model in \empv~places much stronger restrictions on the conditions where stars may form, with more centrally concentrated star formation, thus leading to (a sub-set of) much smoother, more symmetric \hi discs.

While non-parametric morphological indicators have been measured for (some) THINGS and WHISP galaxies, a direct comparison between observations and our simulations is not straightforward. The absolute values of the non-parametric indicators depend sensitively on the maps from which they are calculated (and can be biased by the observational S/N and point-spread function of the beam, \citealt{Thorp2021}). Thus a comparison between \hi mock observations of the simulations and observations, measuring the non-parametric morphological indicators in exactly the same way would be desirable. We defer this to future work.

\section{Asymmetry as tracer of the baryonic physics}\label{s:rfr}
We perform a random forest regression (RFR), using \texttt{scikit-learn} \citep{scikit-learn}, to gain a better intuition about the drivers behind the different \hi morphologies across the simulated galaxies. The random forest model is trained to predict each non-parametric morphological indicator based on a number of global and local galaxy properties. It returns a list of feature importances, i.e.\ how important each respective property is for the model to predict Gini, Smoothness or Asymmetry. RFR is one of the most powerful and simple non-linear regression methods, particularly suited to asses feature importances for large astronomical datasets \citep{Donalek2013}. Hence, we turn to RFRs to infer which features are the most important to predict the non-parametric morphological indicators, because there are a large number of plausible potential drivers, many of which show some level of (anti-)correlation with the non-parametric morphological indicators. Specifically, we consider the global parameters galaxy stellar mass, $\Mstar$, central stellar surface density, $\mus$, the stellar velocity dispersion in the central kpc $\sigma_{\ast, \rm central}$, and the radius of the galaxy $\Rg$ (= 0.1$R_{\rm halo}$); the HI and gas properties such as the \hi mass enclosed in the \hi scale radius, $\MHI (R<R_{\rm HI})$, the \hi scale radius $R_{\rm HI}$, the \hi mass enclosed within the galaxy $\MHI (R<\Rg)$, the total gas mass enclosed within the galaxy $\Mgas (R<\Rg)$, the average \hi and total gas surface densities within the central kpc, $R_{\rm HI}$ and the galaxy, $\SHI (R<1~\kpc)$, $\Sg$, $\SHI (R<R_{\rm HI})$, $\fHI$, ($\fg)$ respectively; the SFR, specific SFR (sSFR = SFR/$\Mstar$), \hi SFE (=SFR/$\MHI$) and star-forming main sequence offset (with respect to the \citealt{Catinella2018} xGASS main sequence) for 10 and 100 Myr averaging timescales and outflow rates for the total gas, $\dot{M}_{\rm gas, out}$, and the HI, $\dot{M}_{\rm HI, out}$, calculated, as in Section~\ref{ss:HIsh}, as the mass flux away from the midplane in a 500 pc thick slab, 5 kpc above and below the galactic midplane. For all physical quantities listed, a single value per galaxy is used. In addition to performing the random forest regression for each different physics sub-set of galaxies, we also perform one for the combined data set including all simulated galaxies, where we add an additional flag to denote the sub-grid physics model\footnote{While the only difference between \empc~and \empv~is in the star formation sub-grid model, this flag does implicitly fold in differences between {\sc{gizmo}} and {\sc{arepo}} and their respective treatment of physical processes other than star formation and stellar feedback, in addition to those differences, for \fc box.} 
of the simulations. 

It is important to stress that the small sample size of 46 galaxies in the combined sample necessarily limits the predictive strength of the RFR, especially when considering the different baryonic physics sub-samples. In those cases, the RFR offers an overview of features that could be important, and for which we have verified that the non-parametric morphological indicator shows a statistically significant, monotonic (anti-) correlation with the respective quantity using the Spearman rank correlation coefficient. To get a better handle on the uncertainty associated with the RFR, we determine the mean feature importances and their standard deviation via bootstrapping. Specifically, we generate 30 different train/test sets with the \texttt{train\_test\_split} function (using a \texttt{test\_size} of 0.2) in \texttt{scikit-learn} by varying the seed of the random number generator, which governs the division into train and test sample. Subsequently, we run 30 RFRs with different random number seeds on each training set, resulting in a total of 900 RFRs. 

\begin{figure*}
    \centering
    \includegraphics[width=\linewidth]{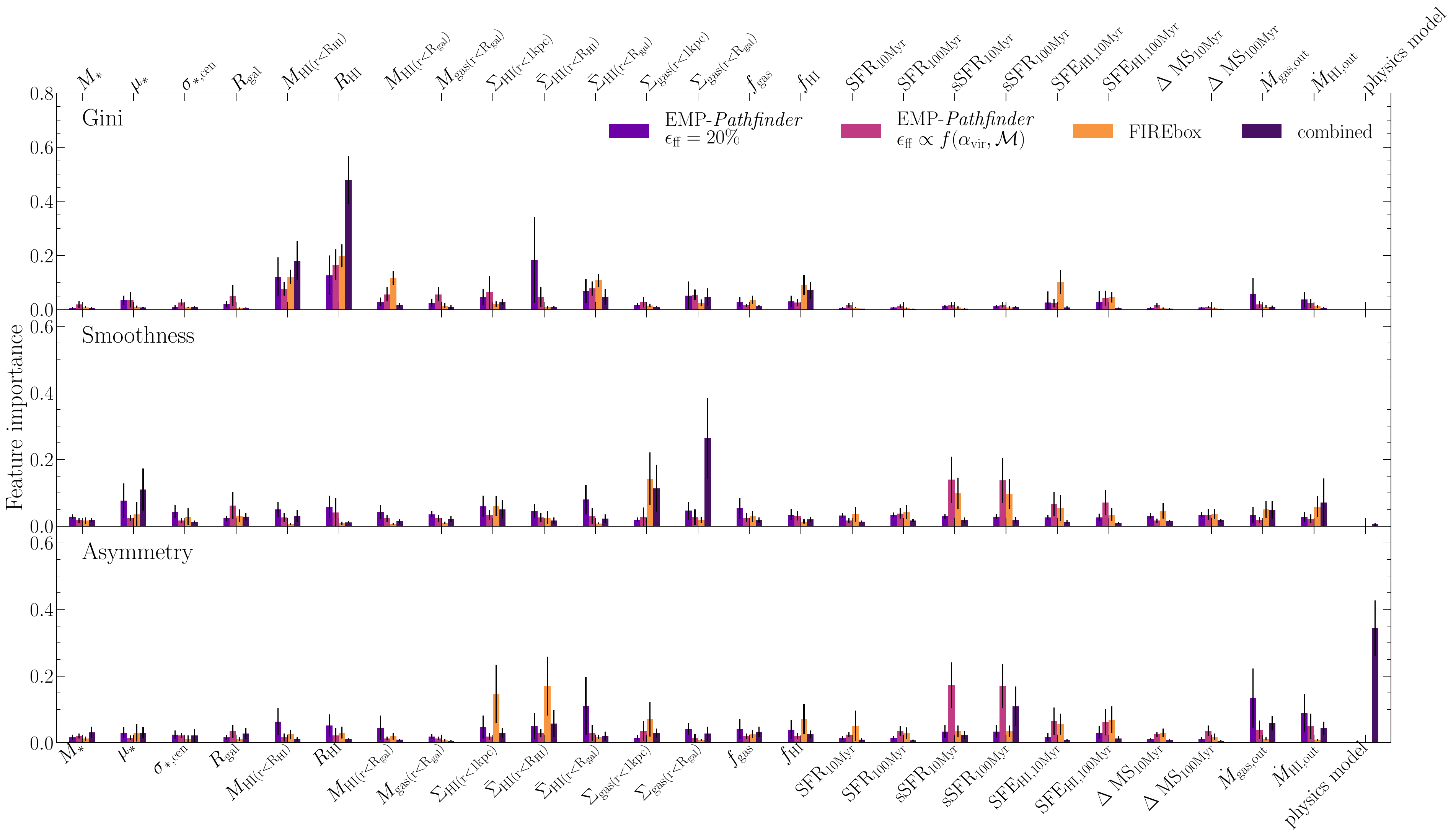}
    \caption{Results of the random forest regression for the Gini, Smoothness and Asymmetry parameters (top to bottom). Feature importance are shown in purple, magenta, orange and dark blue for the constant SFE \emp, multi free-fall \emp, \fc box and the combined sample respectively. Bars show the mean feature importance, with error bars denoting the standard deviation, determined via bootstrapping of 30 random forest regressions on each of the 30 randomly selected sub-samples. The most important feature to predict the Asymmetry of \hi discs in the combined data set is the sub-grid physics with which the galaxies were simulated.}
    \label{fig:rrf}
\end{figure*}

Figure~\ref{fig:rrf} shows the results of the random forest regression for each sub-set, as well as the combined sample of simulated galaxies. For Gini, the most relevant features are $R_{\rm HI}$ and the mass of HI within $R_{\rm HI}$ and $\Rg$ 
, which are strongly correlated (Figure~\ref{fig:HIMD}). Since Gini measures the inequality in the density distribution, this is not surprising considering the median \hi surface density profiles (Figure~\ref{fig:HISD}) of the gas discs. Especially for \empv~and \fc box galaxies the median \hi surface densities vary little in the central region before exponentially declining at $R\geq 0.8 R_{\rm HI}$, which in combination with the density cuts leads to a sharp contrast between high surface density pixels and those without. The median \hi surface density of the \empc~galaxies already declines gradually from the centre of the disc, which plausibly accounts for the feature importance of the average \hi surface density within $R_{\rm HI}$ in predicting Gini for this sub-set of galaxies. 

Considering the combined data-set, the most important features to predict the Smoothness are the average (central) total gas surface density and the central stellar surface density. These properties all show a mild anti-correlation with Smoothness, i.e.\ the denser the gas and the deeper the potential, the smoother the \hi disc. Interestingly, this matches the findings of \citet{Davis2022} for the most important predictors for the Smoothness of the central molecular gas discs in a sample of local early- and late-type galaxies. In \citet{Davis2022} this is interpreted as a consequence of the dynamical suppression of fragmentation, where shear from the deeper central gravitational potential in high-$\mus$ galaxies inhibits fragmentation (see also \citealt{Gensior2020,Gensior2023b}). The primary dependence on the central molecular gas surface density is also a consequence of the stars dominating the potential and gravitational stability \citep{Davis2022}. Our sample consists of star-forming, mostly disc-like galaxies, and the \hi discs extend to 10s of kpc from the centre of the galaxy. It is therefore more likely that the Smoothness depends on these quanties (especially in the \fc box and \empc~cases), because these quantities determine where stars form and how effectively outflows can be driven and thus how porous the ISM becomes. For \empv~galaxies, the specific star formation rate averaged over both 10 and 100 Myr is selected as the most important feature, which shows a correlation with the Smoothness (i.e.\ galaxies that form more stars with respect to their total stellar mass have a less smooth \hi disc). This is likely because in this sub-grid star formation model the SFR is less dependent on the gas density, because the $\eff$ is set by the local turbulent properties instead. It will thus depend somewhat on the potential, as galaxies with higher-$\mus$ are likely to have a higher turbulent velocity dispersion (in the central regions) thereby inhibiting star formation somewhat. The dependence of the Smoothness of the \fc box galaxies on the sSFR is likely the result of the strong primary dependence on the gas surface density, which governs the star formation activity in \fc. 

Asymmetry is where the feature importances between the individual galaxy sub-sets and for the combined data-set differ the most. The most important feature to predict the Asymmetry of \hi discs for the combined set is the baryonic physics model (0.34$\pm$ 0.08), highlighting that it is very sensitive to the subtle differences in star formation and feedback physics (and their consequences). This suggests, therefore, that the Asymmetry of \hi discs could be a very promising observable to further our understanding of the baryonic physics at play in our Universe. The most important features for \empc~galaxies are the outflow rates. They are correlated with the Asymmetry, i.e.\ stronger outflows lead to more asymmetric discs. For the \empv~galaxies, the sSFR is the most important, again showing a correlation with the Asymmetry. The (central) \hi density is the most important feature for \fc box galaxies, which is anti-correlated with the Asymmetry. This is perhaps counter-intuitive, as a higher surface density should lead to higher SFRs and thus more feedback. However, denser gas will be able to cool more efficiently and is more strongly gravitationally bound, thus making it harder to disturb. In short, the most important features to predict the Asymmetry of the \hi discs all directly relate to the different physics models, where stars form and how effective the stellar feedback will be in affecting the gas distribution. 

\begin{figure*}
    \centering
    \includegraphics[width=\linewidth]{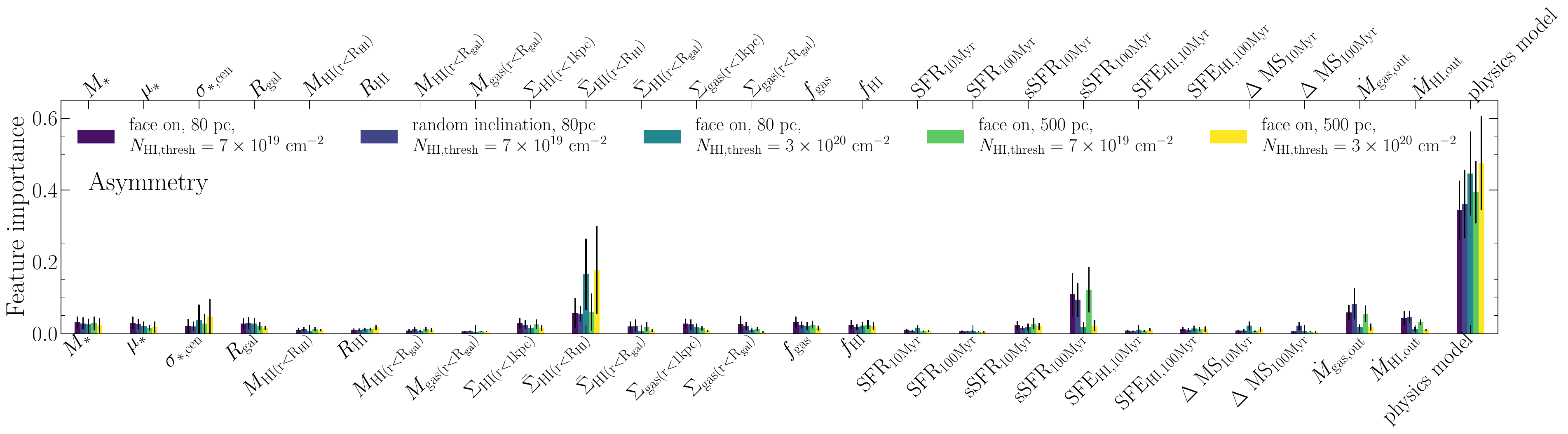}
    \caption{Results of the random forest regression trained on the combined data set, to identify the best predictor for the Asymmetry of the \hi discs. Coloured bars and error bars show the mean and standard deviation determined via bootstrapping of performing 30 random forest regressions on each of the 30 randomly selected sub-samples. Different colours indicate the results of the random forest regression for the random inclination, different resolution and density threshold maps from which the non-parametric morphologies were determined. The physics model is selected as the most important feature in all cases and by a large margin.}
    \label{fig:rfrA_dm}
\end{figure*}

We repeat the RFR for the non-parametric indicators computed from maps at coarser resolution, with a higher density threshold and random inclinations $<70^{\circ}$. Figure~\ref{fig:rfrA_dm} shows the feature importances for predicting the Asymmetry of the combined dataset. Crucially, the physics model remains the most important feature (feature importances 0.34 $\pm$ 0.08 {--} 0.48 $\pm$ 0.12) by far. The only qualitative difference between the feature importances for the different maps is the selection of average global surface density as the second most important feature (compared to the sSFR) for the high-density threshold maps. This can be understood as the morphologies of galaxies with lower surface densities are more strongly affected by the stricter cut. 

\begin{figure*}
    \centering
    \includegraphics[width=\linewidth]{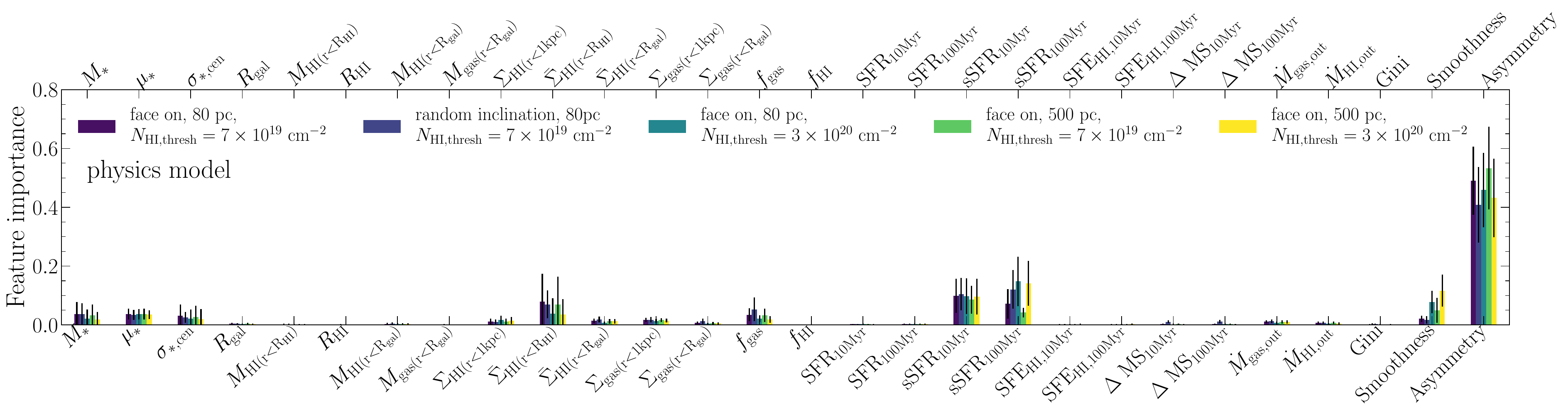}
    \caption{Results of the random forest regression to identify the best predictor for the baryonic physics model with which the galaxies were simulated. Coloured bars and error bars show the mean and standard deviation determined via bootstrapping of performing 30 random forest regressions on each of the 30 randomly selected sub-samples. Different colours indicate the results of the random forest regression for the random inclination, different resolution and density threshold maps from which the non-parametric morphologies were determined. The Asymmetry is selected as the most important feature in all cases and by a large margin.}
    \label{fig:rfr_rev}
\end{figure*}

To further validate our results of the strong dependence of the \hi disc Asymmetry on the underlying baryonic physics model, we also perform the reverse test. We use the combined dataset, include Gini, Asymmetry and Smoothness in the list of potential features and perform a RFR to predict the (flag for the) baryonic physics model of the simulated galaxies. The results for this RFR are shown in Figure~\ref{fig:rfr_rev}. In all cases, the Asymmetry is by far the most important feature to predict the baryonic physics model (feature importance in the range 0.41$\pm$0.12 {--} 0.53$\pm$0.13), independent of the inclination, resolution, or density threshold of the maps on which the non-parametric morphological indicators were calculated. This lends further support to the Asymmetry being a promising observable in order to validate or disprove physics models by comparison to observations.

Although limited by the small numbers in our sample, this approach demonstrates the power of combining complex statistical models (machine learning) with simulations run with different baryonic physics models to learn more about their effects and highlight particularly promising observables. A similar approach was taken by \citet{Maccio2022} to compare galaxies from the Numerical Investigation of a Hundred Astrophysical Objects (NIHAO; \citealt{Wang2015}) suite, simulated with the same constant SFE but different density thresholds for star formation, to THINGS galaxies to establish which gas maps best match the observations. They used a deep neural network, trained on moon craters, to identify features on simulated and observed gas maps. They found that the number of features in maps from galaxies simulated with a density threshold of 80 atoms per cubic centimetre (the highest in their sample) yields the best agreement with THINGS galaxies.
Since NIHAO also includes early stellar feedback and a cold ISM, it would be interesting to see where these galaxies fall in the Smoothness-Asymmetry parameter space discussed in Section~\ref{s:HImorph} and here. Care must be taken when comparing with (more) galaxies simulated with different hydro-dynamical codes, as even the gas disc morphologies of individual galaxies simulated with very `similar' sub-grid physics appear visually different, with the largest differences between Eulerian and Langrangian codes \citep[e.g.][]{Kim2016,Hu2023}. However, the results presented in this paper are in large part driven by differences between the two \emp~sub-sets of galaxies, where numerics-wise everything beyond the differences in SFE is exactly the same, and should therefore be robust. None the less, the possibility that some of the differences between \fc box and \emp~stem from subtle differences in the numerics between {\sc{gizmo}} and {\sc{arepo}}, rather than the differences in baryonic physics, cannot be excluded.

\section{Concluding remarks}\label{s:conclusion}
In this paper, we analysed the properties of 46 Milky Way halo-mass galaxies from the \emp~suite of cosmological zoom-in simulations and the \fc box cosmological volume at $z=0$, with a focus on their \hi disc properties. The sample comprises galaxies simulated with three different star formation models. This is because the \emp~galaxies were evolved to $z=0$ once with a constant $\eff=20$~per~cent and once with an environmentally-dependent, turbulence-based $\eff$; the \fc-2 model also features a constant $\eff=100$~per~cent, but additionally requires the gas to be locally self-gravitating, Jeans unstable and molecular. All simulations include supernovae feedback and stellar winds from evolved stars, but only \fc box includes a contribution from early stellar feedback i.e.\ the winds, radiation pressure, winds, photoelectric heating and photoionization by massive OB stars. Therefore, this set of simulations is ideally suited to assess the impact of small differences in the sub-grid baryonic physics on galaxy properties after self-consistent evolution across cosmic time. Our results can be summarised as follows. 

\begin{enumerate}[leftmargin=0.5cm]
    \item Despite the similar halo masses, the galaxy stellar masses differ between the different baryonic physics sub-sets, though they are generally consistent with the large scatter of the stellar-to-halo mass relation; half of the \empc~galaxies have lower and a quarter of the \fc box galaxies have higher stellar masses compared to that of the Milky Way. Nonetheless, the \hi-to-stellar mass fractions of each sub-set are consistent with those of the star-forming xGASS \citep{Catinella2018} galaxies. Furthermore, all galaxies lie on the \hi mass-size relation in good agreement with observations. 
    
    \item The median \hi surface density profiles differ in both shape and normalisation within the inner part of the \hi disc ($<0.8R_{\rm HI}$), before following the characteristic exponential decline observed in the \hi discs of local late-type galaxies. 
    
    \item The \hi discs of \empc~galaxies are consistently thicker compared to the \empv~galaxies, to \fc box galaxies, and to the observed THINGS galaxies \citep{Bacchini2019a}, highlighting that a cold ISM might be a necessary, but not a sufficient condition to reproduce the thin gas discs observed in the local Universe. 
    
    \item The \hi disc morphologies, as quantified by the non-parametric morphological indicators Smoothness and Asymmetry, differ significantly between the different galaxy sub-sets. The \empv~galaxies have the smoothest and most rotationally symmetric \hi discs, the \empc~simulations the most asymmetric discs with large low-density areas, while the \fc box discs have very similar morphologies (narrow range in Asymmetry and Smoothness), with Asymmetries between those of the \empv~and \empc~galaxies. 

    \item Random forest regression selects the baryonic physics model as the most important feature to predict the Asymmetry of the \hi discs, and vice-versa the Asymmetry (followed by the specific star formation rate) is the most important feature for predicting the baryonic physics model. This further highlights how sensitive the gas disc morphologies are to subtle differences in the baryonic physics and suggests their Asymmetry could be a powerful observable to better constrain the physics at play in the local Universe. 
\end{enumerate}

Our work demonstrates the sensitivity of gas disc morphologies to underlying star formation and feedback physics. However, to gain a more complete picture of their impact on galaxy evolution, a more thorough exploration including an entire range of halo masses, using more simulations with different physics, is required. For example, despite the different stellar feedback in \emp~and \fc box, the implementation of the feedback injection is similar, as the \emp~implementation follows the \fc-2 \citep{Hopkins2018} mechanical feedback coupling algorithm. Thus, it would be informative to see how different implementations, such as blastwave feedback \citep[e.g.][]{Stinson2010} or kinetic feedback \citep[e.g. AURIGA;][]{Grand2017}, affect the distributions of Asymmetry and Smoothness. Furthermore, since our results remain robust even for coarse resolutions of 500 pc, this potentially enables an exploration of how the \hi disc morphology of galaxies is affected by AGN feedback, by drawing on cosmological simulations that also include AGN feedback (e.g.\ EAGLE and IllustrisTNG, which have a resolution of several hundred pc). Complementarily, a large sample of high-resolution observations of \hi disc morphologies from the SKA and its precursors (e.g.\ MHONGOOSE) would help validate or disprove these different physics models.

\section*{Acknowledgements}
We thank an anonymous referee for their helpful and constructive report. We thank Kai Polsterer for his input on machine learning methods and feature importance. JG gratefully acknowledges financial support from the Swiss National Science Foundation (grant no CRSII5\_193826). 
RF acknowledges financial support from the Swiss National Science Foundation (grant nos. 194814, 200021\_188552). 
MRC gratefully acknowledges the Canadian Institute for Theoretical Astrophysics (CITA) National Fellowship for partial support.  This work was supported by the Natural Sciences and Engineering Research Council of Canada (NSERC). 
STG gratefully acknowledges the generous and invaluable support of the Klaus Tschira Foundation, in addition to funding from the European Research Council (ERC) under the European Union’s Horizon 2020 research and innovation programme via the ERC Starting Grant MUSTANG (grant agreement number 714907).
AW received support from: NSF via CAREER award AST-2045928 and grant AST-2107772; NASA ATP grant 80NSSC20K0513; HST grants AR-15809, GO-15902, GO-16273 from STScI.
JMDK gratefully acknowledges funding from the DFG through an Emmy Noether Research Group (grant number KR4801/1-1) and from the European Research Council (ERC) under the European Union's Horizon 2020 research and innovation programme via the ERC Starting Grant MUSTANG (grant agreement number 714907). COOL Research DAO is a Decentralised Autonomous Organisation supporting research in astrophysics aimed at uncovering our cosmic origins.
Support for PFH was provided by NSF Research Grants 1911233, 20009234, 2108318, NSF CAREER grant 1455342, NASA grants 80NSSC18K0562, HST-AR-15800.
JM is funded by the Hirsch Foundation.

The \emp~production runs were run on the Graham supercomputing cluster from Compute Ontario, and on the BinAc cluster from the University of T{\"u}bingen. The research was enabled in part by support provided by Compute Ontario (https://www.computeontario.ca) and Compute Canada (www.computecanada.ca). The authors acknowledge support by the High Performance and Cloud Computing Group at the Zentrum f{\"u}r Datenverarbeitung of the University of T{\"u}bingen, the state of Baden-W{\"u}rttemberg through bwHPC and the German Research Foundation (DFG) through grant no. INST 37/935-1 FUGG. We acknowledge PRACE for awarding us access to MareNostrum at the Barcelona Supercomputing Center (BSC), Spain. This work was supported in part by a grant from the Swiss National Supercomputing Centre (CSCS) under project IDs s697 and s698. We acknowledge access to Piz Daint at the Swiss National Supercomputing Centre, Switzerland under the University of Zurich's share with the project ID uzh18. This work made use of infrastructure services provided by S3IT (www.s3it.uzh.ch), the Service and Support for Science IT team at the University of Zurich.

\textit{Software:} This work made use of the \texttt{python} packages \texttt{h5py} \citep{h5py}, \texttt{jupyter-notebooks} \citep{Kluyver2016jupyter}, \texttt{matplotlib} \citep{matplotlib}, \texttt{NumPy} \citep{numpy}, \texttt{pandas} \citep{mckinney-proc-scipy-2010,reback2020pandas}, \texttt{scikit-learn} \citep{scikit-learn}, and \texttt{SciPy} \citep{2020SciPy-NMeth}.  

\section*{Data Availability}
The data supporting the plots in this article will be shared on reasonable request to the corresponding author.



\bibliographystyle{mnras}
\bibliography{references} 




\appendix

\section{Comparing different models for obtaining HI and H$_2$ fractions from the neutral gas}\label{A:H2conv}
\begin{figure*}
    \centering
    \begin{subfigure}[b]{0.45\linewidth}
    \includegraphics[width=0.95\linewidth]{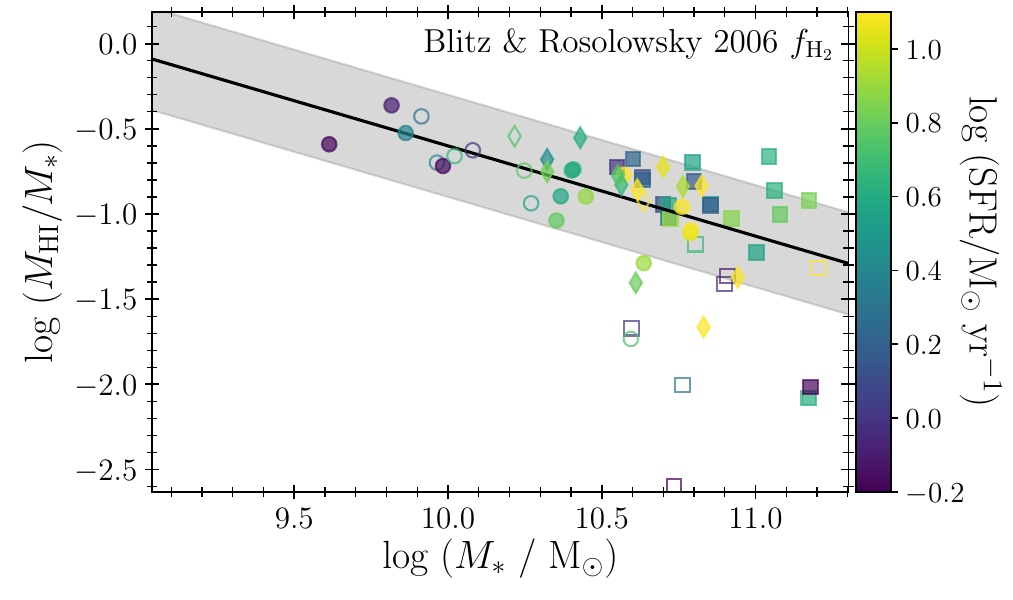}
    \end{subfigure}
    \begin{subfigure}[b]{0.45\linewidth}
    \includegraphics[width=0.95\linewidth]{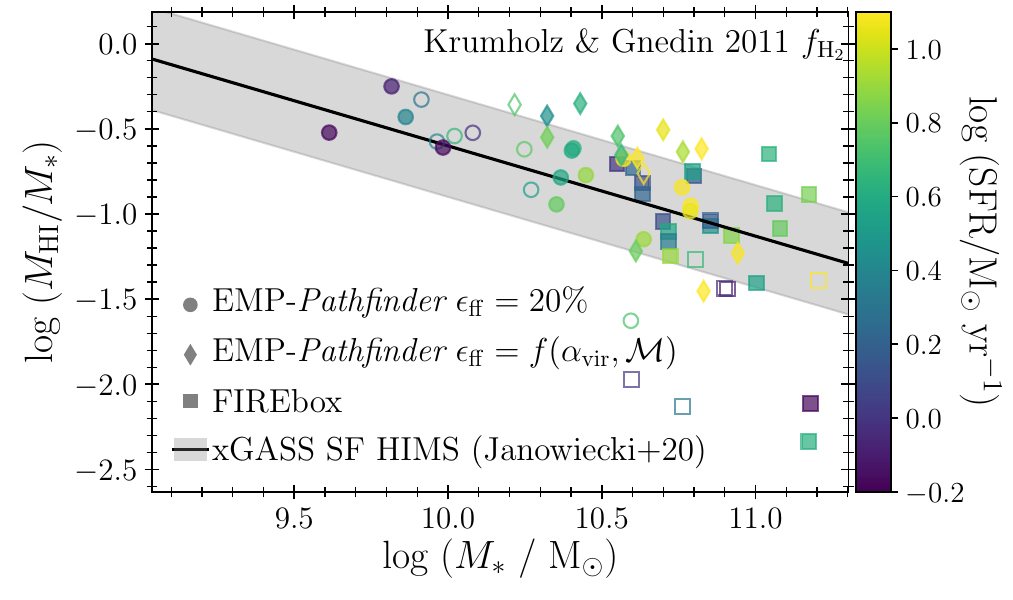}        
    \end{subfigure}
    \caption{HI-to-stellar mass fraction as a function of stellar mass of the simulated galaxies in our sample. The left and right panels show the simulation data with the \hi fraction calculated using the pressure-based \citet{Blitz2006} model and the \citet{Krumholz2011} formalism, respectively. The data are colour-coded by their star formation rate, with different symbols denoting the different sub-grid physics samples. Empty symbols denote the galaxies excluded from further analysis, due to interactions. The data are overplotted on the \hi main sequence relation of the star-forming xGASS \citep{Catinella2018} galaxies from \citet{Janowiecki2020} shown as a solid black line, and the 0.3 dex uncertainty in grey shading. The global \hi content of galaxies is larger in case of the non-fiducial model. \empv~is affected the most, with shifts of $\sim$0.3 dex, moving 6 (7) galaxies above the 0.3 dex scatter on the xGASS \hi main sequence when calculating the \hi fraction using \citet{Krumholz2011}.}
    \label{fig:AH2_HIMS}
\end{figure*}

As discussed in Section~\ref{ss:calcfHI}, we use two different models to infer the `true' \hi content from the neutral gas fraction in the \emp~and \fc box galaxies. The \hi fraction of \emp~galaxies is calculated using the empirical, pressure-based model of \citet{Blitz2006}. For \fc box galaxies, we use the \citet{Krumholz2011} model, based on idealised cloud simulations, which is also used to estimate the H$_2$ fraction at run time in \fc. The use of two different models is motivated by attempting to match the observed density threshold of $\sim$10 $\Msun~\pc^{-2}$ at which $\SHI$ saturates and above which gas becomes molecular \citep[e.g.][]{Blitz2006,Bigiel2008,Leroy2008}. Calculating the \hi fraction with the respective non-fiducial model leads to higher \hi abundances that result in surface densities in excess of 20 $\Msun~\pc^{-2}$ for many galaxies in the respective sub-set. This is particularly visible for \fc box galaxies in the left panel of Figure~\ref{fig:AH2_HISD}. In this Appendix, we show versions of Figures~\ref{fig:HIMS} {--} \ref{fig:hHI} and Figure~\ref{fig:HImorph} where $\fHI$ has been calculated with the same model for all simulated galaxies for both the \citet{Blitz2006} and \citet{Krumholz2011} model in Figures~\ref{fig:AH2_HIMS} {--} \ref{fig:AH2_GAS}.

\begin{figure*}
    \centering
    \begin{subfigure}[b]{0.45\linewidth}
    \includegraphics[width=0.95\linewidth]{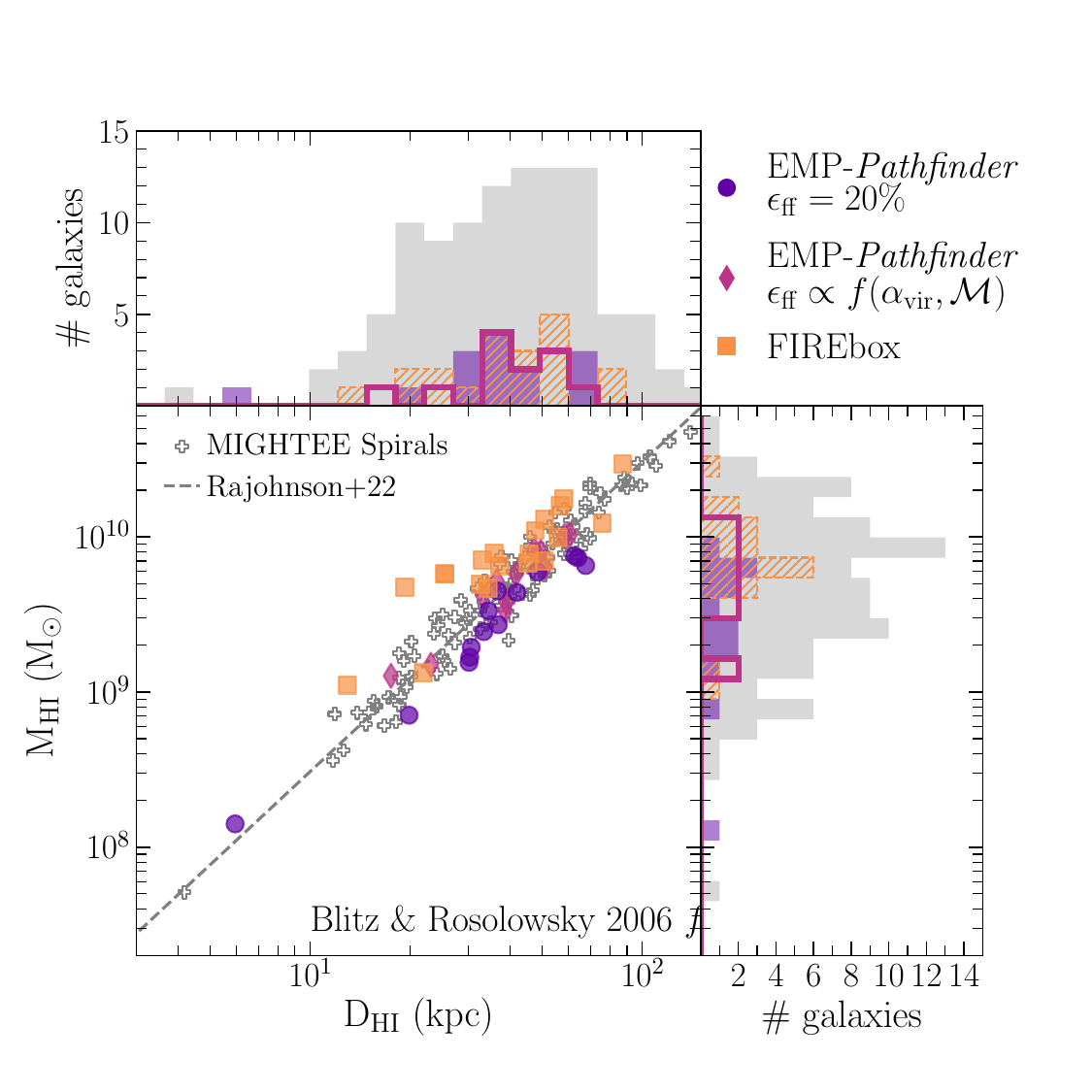}
    \end{subfigure}
    \begin{subfigure}[b]{0.45\linewidth}
    \includegraphics[width=0.95\linewidth]{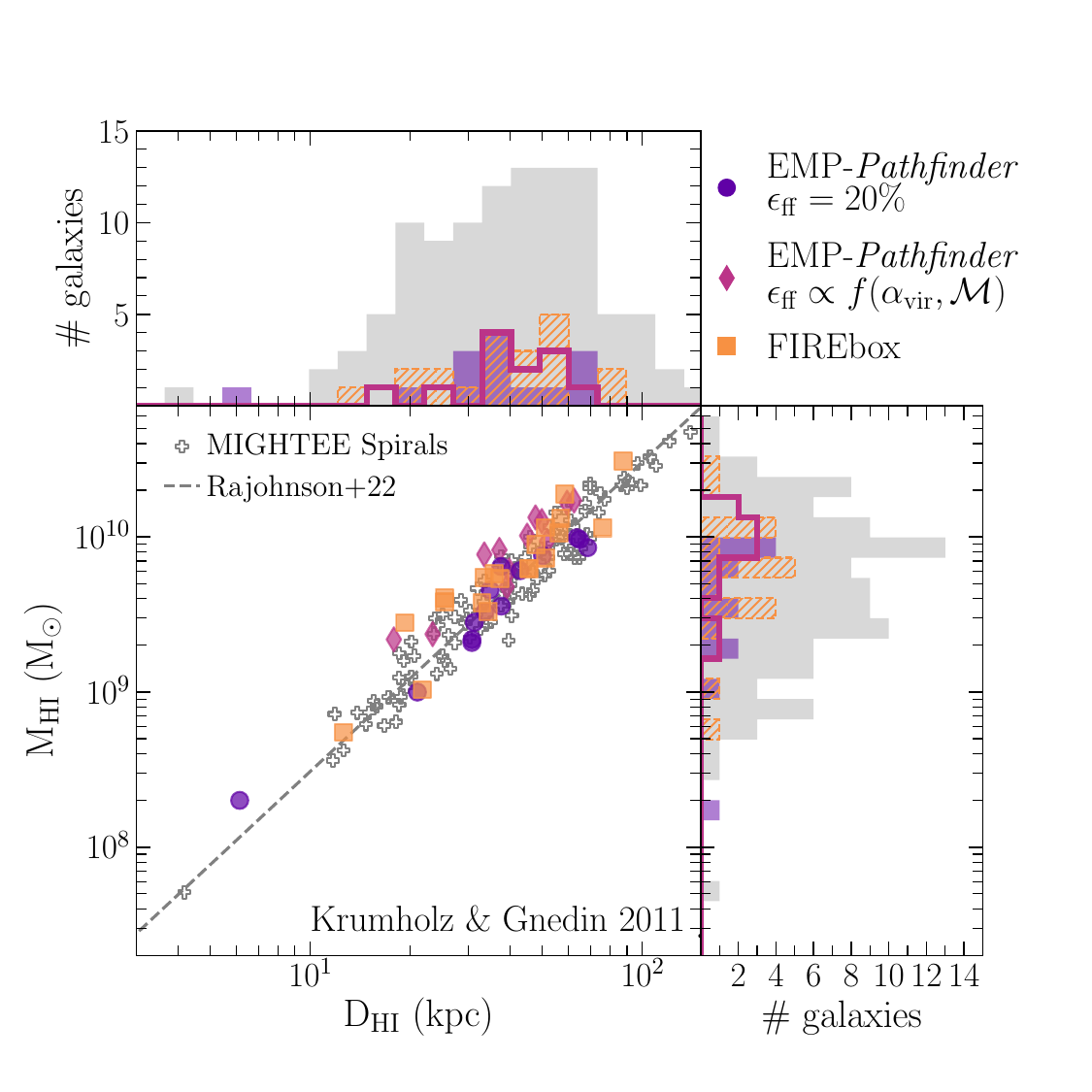}        
    \end{subfigure}
    \caption{\hi mass-size relation for the galaxies in our sample. The left and right panels show the simulation data with the \hi fraction calculated using the pressure-based \citet{Blitz2006} model and the \citet{Krumholz2011} formalism, respectively. Purple circles denote the \empc~galaxies, magenta diamonds denote the \empv~galaxies, and orange squares denote the \fc box galaxies. Histograms show the marginal distributions for each indicator and galaxy sub-set. The grey dashed line shows the HI mass-size relation from MIGHTEE \citep{Rajohnson2022}. Galaxies tend to be more offset from the \hi mass-size relation with the non-fiducial model, which is especially pronounced for the smaller \fc box discs and the \empv~galaxies.}
    \label{fig:AH2_HIMD}
\end{figure*}

\begin{figure*}
    \centering
    \begin{subfigure}[b]{0.45\linewidth}
    \includegraphics[width=0.95\linewidth]{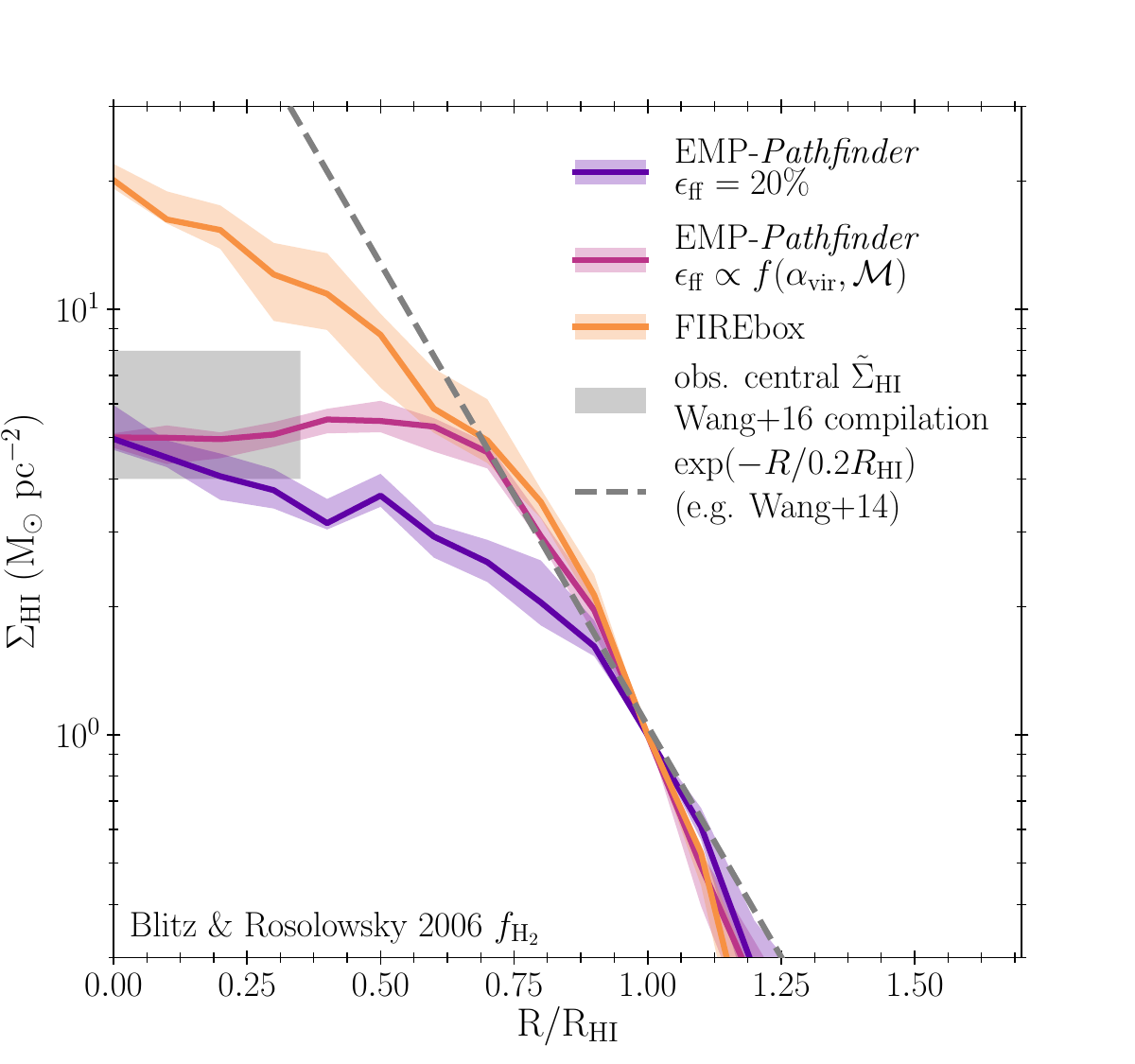}
    \end{subfigure}
    \begin{subfigure}[b]{0.45\linewidth}
    \includegraphics[width=0.95\linewidth]{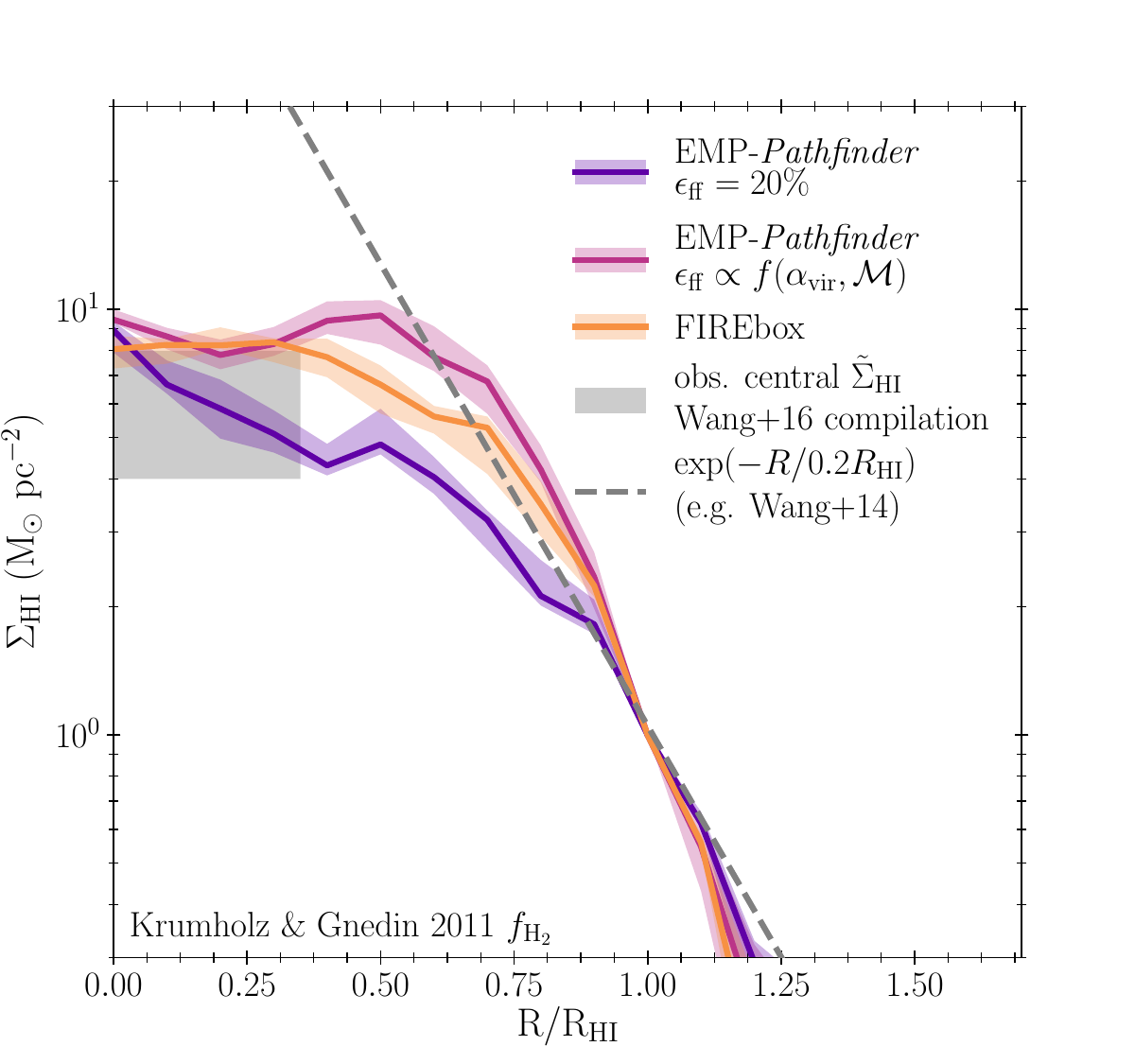}        
    \end{subfigure}
    \caption{Radial \hi gas surface density profiles. The left and right panels show the simulation data with the \hi fraction calculated using the pressure-based \citet{Blitz2006} model and the \citet{Krumholz2011} formalism, respectively. Coloured lines and shaded regions denote the median and the error on the median, determined via bootstrapping, of each different baryonic physics sub-sample, respectively, with the \citet{Wang2014} exponential fit to the outer regions of observed HI surface density profiles overplotted as a grey-dashed line. For comparison, the grey shaded box indicates the central range of the median $\SHI$ from 8 different observational samples, compiled by \citet{Wang2016}. The median central surface densities reached by the simulated galaxy samples are a factor of $\gtrsim$2 larger for the non-fiducial model, this is particularly pronounced for the \fc box galaxies with a median $\SHI$ of 20 $\Msun~\pc^{-2}$. When using the \citet{Blitz2006} model, the median \hi surface density radial profile of \fc box galaxies continuously declines as a function of radius, as opposed to remaining approximately constant to 0.3$R/R_{\rm HI}$, before declining with the fiducial \citet{Krumholz2011} model.}
    \label{fig:AH2_HISD}
\end{figure*}

\begin{figure*}
    \centering
    \begin{subfigure}[b]{0.495\linewidth}
    \includegraphics[width=\linewidth]{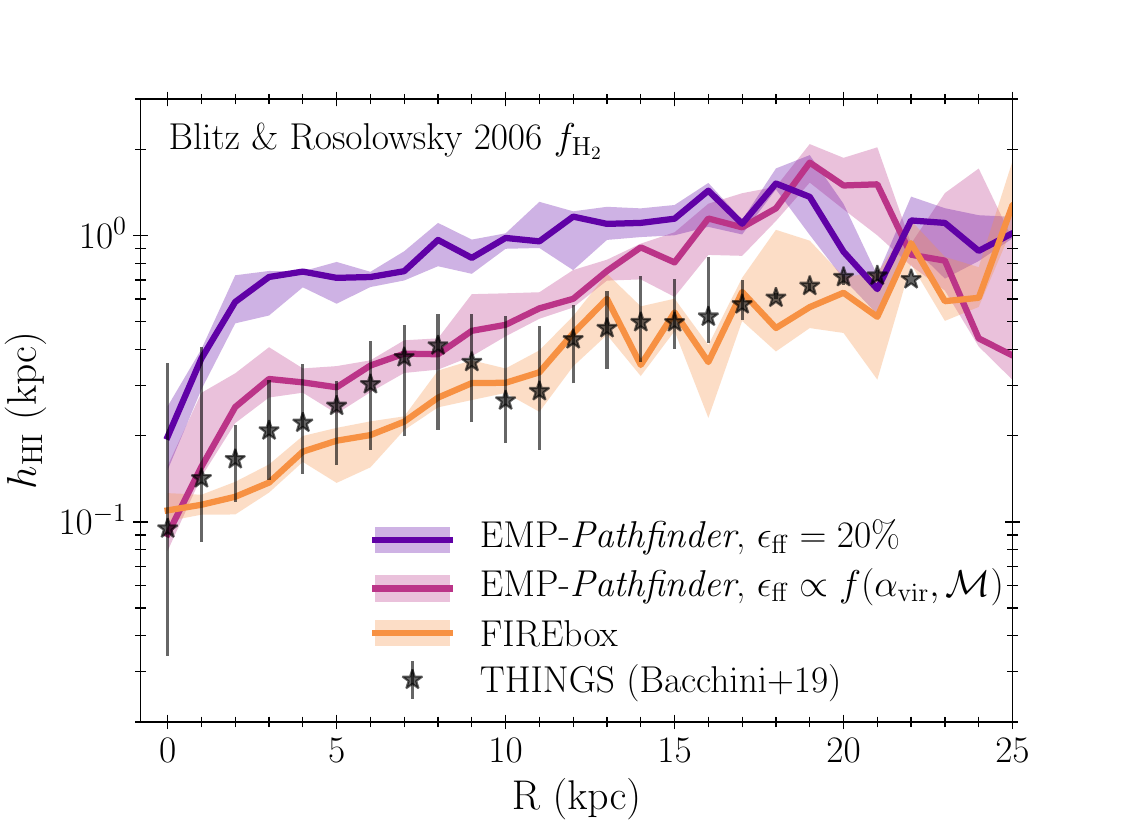}
    \end{subfigure}
    \begin{subfigure}[b]{0.495\linewidth}
    \includegraphics[width=\linewidth]{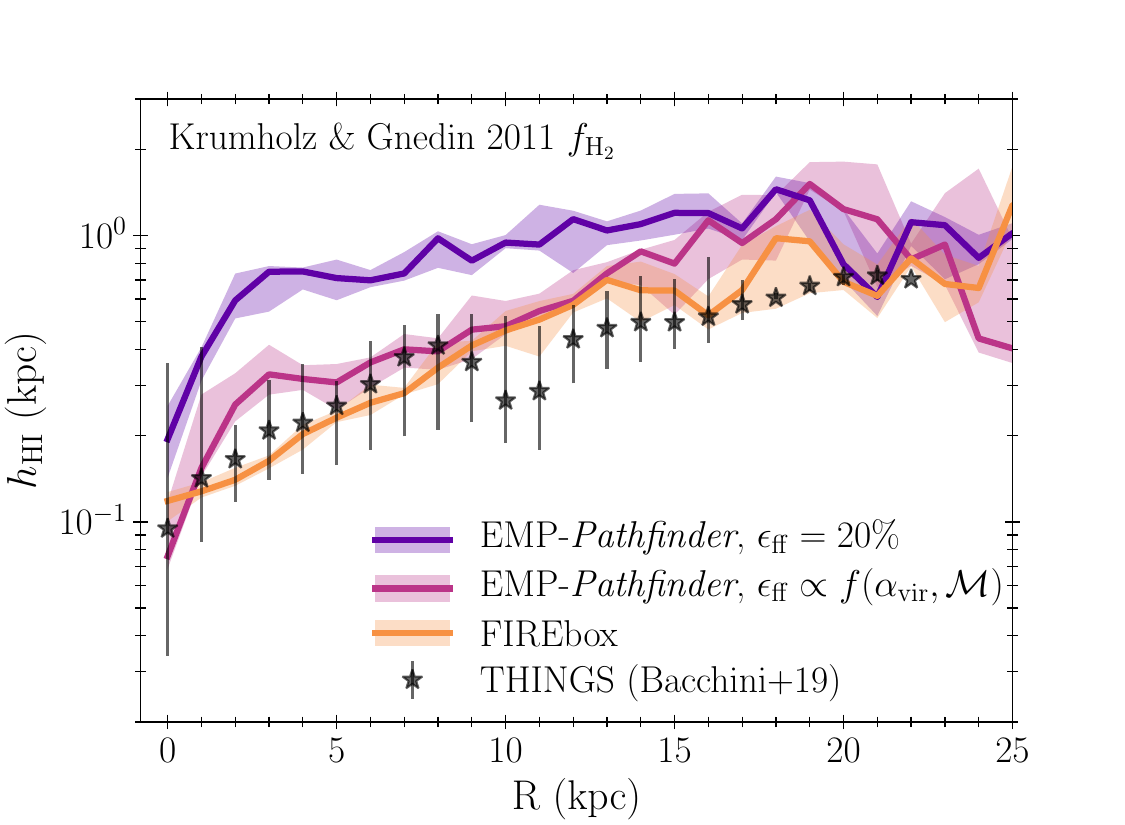}        
    \end{subfigure}
\caption{Radial profiles of the \hi disc scale heights of the simulated galaxies, measured by fitting a Gaussian to the vertical volume density distribution in kpc bins. The left and right panels show the simulation data with the \hi fraction calculated using the pressure-based \citet{Blitz2006} model and the \citet{Krumholz2011} formalism, respectively. Coloured lines and shaded regions denote the median and the error on the median, determined via bootstrapping, of each different baryonic physics subsample, respectively, with the THINGS scale heights \citep{Bacchini2019a} overplotted as black stars. The \hi model has a negligible impact on the measured scale heights. They are marginally smaller when measured from the \hi estimate obtained via the non-fiducial model, due to the larger \hi surface densities in that case.}
\label{fig:AH2_hHI}
\end{figure*}

\begin{figure*}
    \centering
    \begin{subfigure}[b]{0.45\linewidth}
    \includegraphics[width=0.95\linewidth]{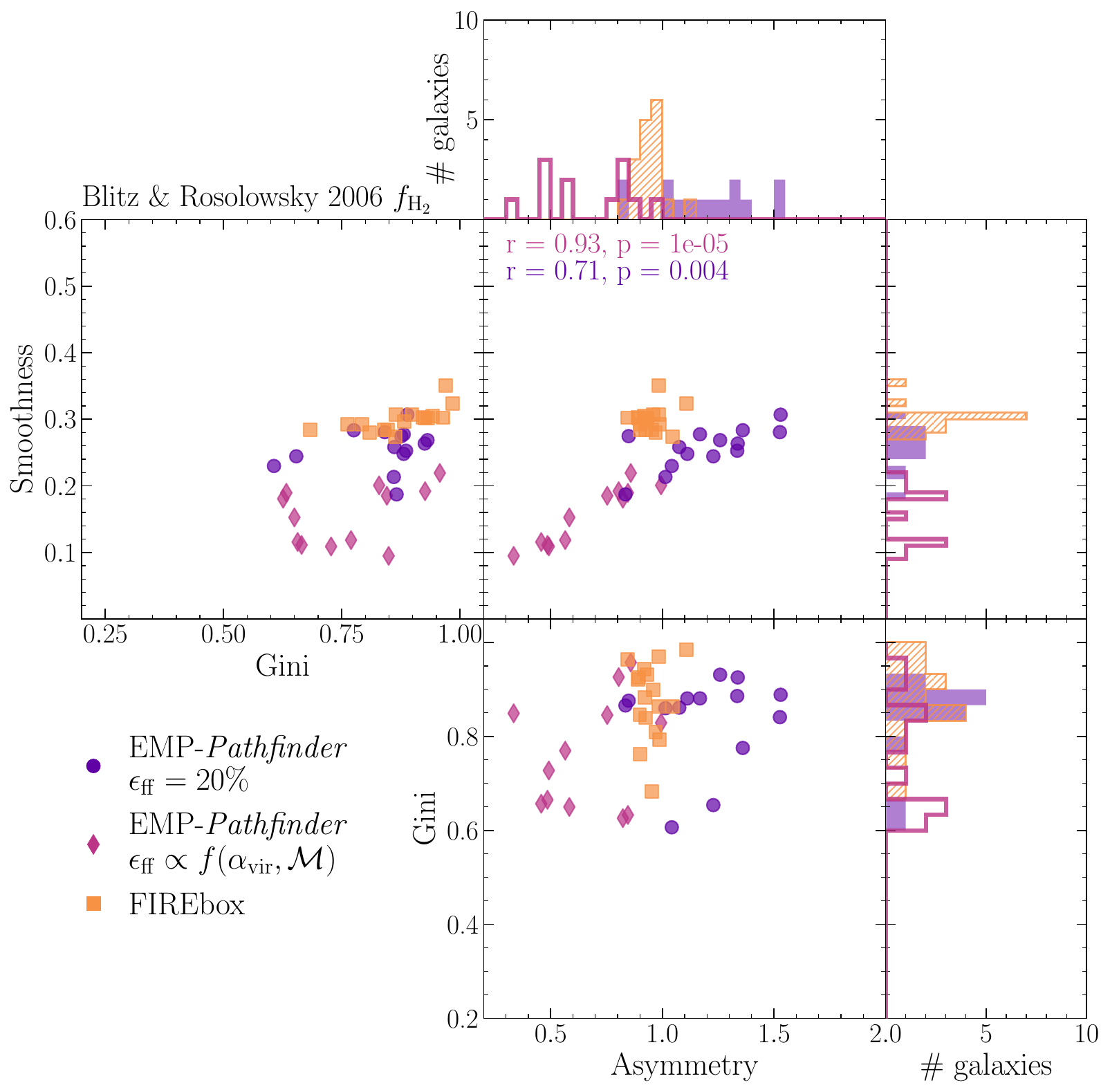}
    \end{subfigure}
    \begin{subfigure}[b]{0.45\linewidth}
    \includegraphics[width=0.95\linewidth]{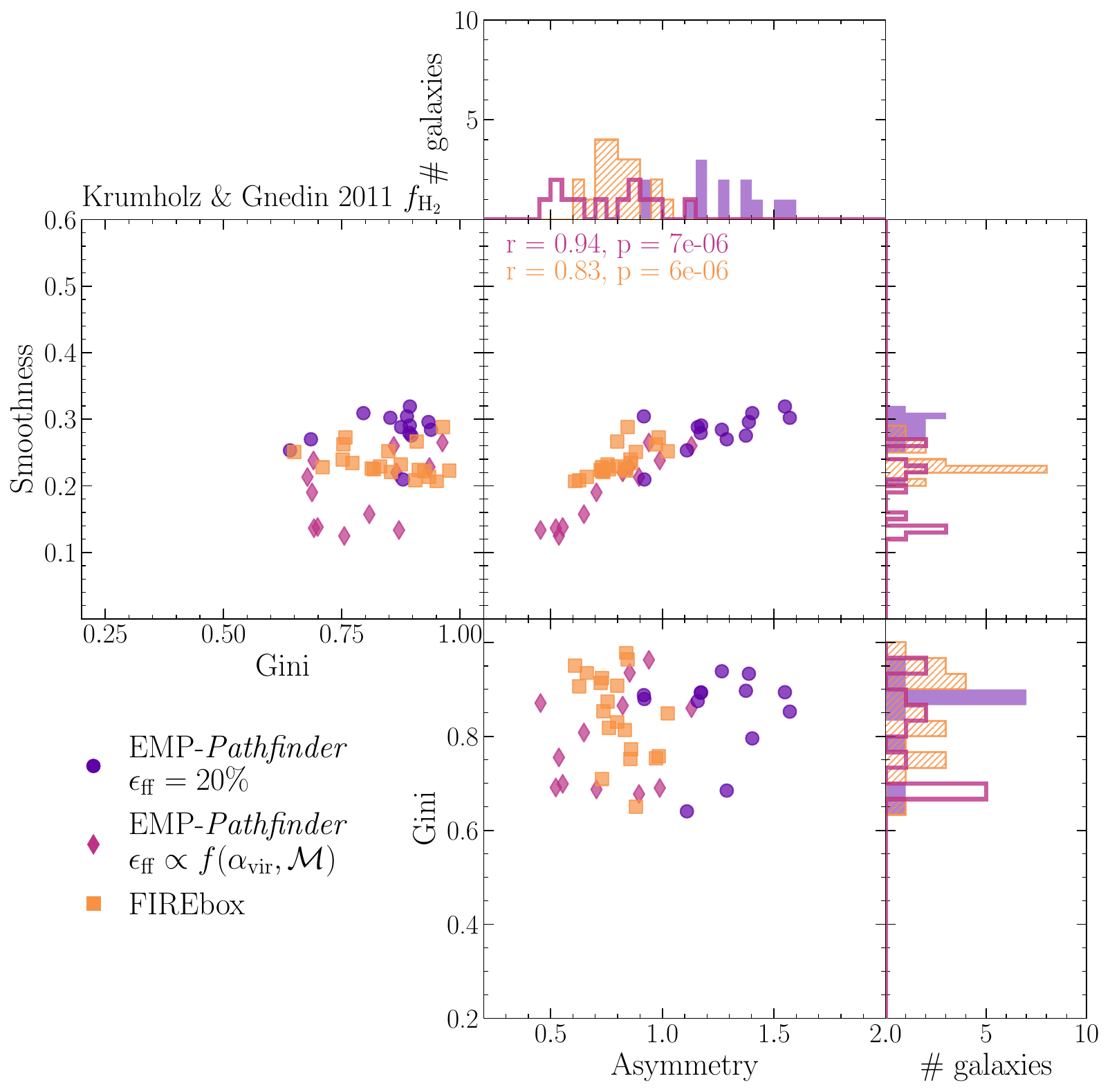}        
    \end{subfigure}
    \caption{\hi morphology as classified by the median non-parametric morphological indicators Smoothness, Gini and Asymmetry (plotted against each other in the three main panels), measured from face-on \hi surface density projections of the galaxies with a resolution of 80 pc and column density threshold of $7\times10^{19}~\rm{cm}^{-2}$. The left and right panels show the simulation data with the \hi fraction calculated using the pressure-based \citet{Blitz2006} model and the \citet{Krumholz2011} formalism, respectively. Purple circles denote the \empc~galaxies, while magenta diamonds denote the \empv~galaxies, and orange squares denote the \fc box galaxies. Histograms show the marginal distributions for each indicator and galaxy sub-set. If a statistically significant correlation between two non-parametric morphology indicators is present, the Spearman rank correlation coefficient and $p$-value are included in the panels, colour indicating which simulated galaxy sub-set they apply to. The overall trends and segregation of the different physics simulated galaxy sub-sets remain the same, irrespective of how the \hi fraction was calculated. The main difference occurs for \fc box galaxies, which are shifted to higher Smoothness and show a much narrower range of Asymmetries when using the \citet{Blitz2006} model.}
    \label{fig:AH2_GAS}
\end{figure*}

The main difference in $\fHI$, as computed from the different models, is an increase in the total \hi mass, i.e.\ shifting to higher HI-to-stellar mass ratios, driven by an increase in central \hi for the non-fiducial model. Therefore, these galaxies are on average more offset from the \hi main sequence (see Figure~\ref{fig:AH2_HIMS}) and the \hi mass-size relation (see Figure~\ref{fig:AH2_HIMD}), because the \hi scale radius, $R_{\rm HI}$ is not sensitive to the enhanced \hi fraction in the central regions. 

Similarly, the median central $\SHI$ of the different physics simulated galaxy sub-sets increases by a factor of 2 (\emp) to 2.5 (\fc box) compared to the fiducial model (Figure~\ref{fig:AH2_HISD}). While the shape of the surface density profile remains similar for the \emp~galaxies, independent of the H$_2$ model, it changes significantly for the \fc box galaxies. With the pressure-based \citet{Blitz2006} model, the median $\SHI$ declines as a function of radius at all radii, contrary to the behaviour with the \fc box-fiducial \citet{Krumholz2011}, where $\SHI$ is approximately constant in the central regions, before declining for $R>0.3R_{\rm HI}$. 

The effect of the different \hi models on calculating the scale height is negligible for the individual galaxy sub-sets (Figure~\ref{fig:AH2_hHI}). In the non-fiducial case, $\hhi$ becomes slightly smaller due to the enhanced \hi density. This results in the \empv~and \fc box median scale heights becoming more offset from each other, albeit still within the range of the THINGS galaxies, when comparing $\hhi$ exclusively calculated via the \citet{Blitz2006} model.

Finally, we show the results of calculating the non-parametric morphological indicators from the two different \hi model gas surface density projections in Figure~\ref{fig:AH2_GAS}. The sub-sets of galaxies simulated with different physics remain clearly separated in the Smoothness-Asymmetry parameter space in both cases, particularly when using the \citet{Blitz2006} model for all galaxies. The non-parametric \hi morphologies of the \fc box galaxies are the most affected by the \hi model, with nearly all galaxies clustered tightly around $A=0.9$ when using the non-fiducial \citet{Blitz2006} model. The \empc~galaxies become a little less smooth (median $S$ shifting from 0.26 to 0.29), and their trend between Smoothness and Asymmetry is no longer statistically significant at the $p<0.01$ level when using the \citet{Krumholz2011} model, while the effect on the \hi morphologies of the \empv~galaxies is negligible. 

In summary, our tests show that while the global \hi mass of the galaxies, and to a lesser extent its distribution within the galaxy, changes depending on the model, our qualitative conclusions remain robust irrespective of the model. Crucially, this is especially the case for the trends of the \hi disc morphologies and the clear separation between the galaxy sub-sets simulated with different physics in the Smoothness-Asymmetry parameter space.

\section{Additional Gini, Smoothness and Asymmetry plots for different density thresholds and resolutions}\label{A:GAS+}

\begin{figure}
    \centering
    \includegraphics[width=\linewidth]{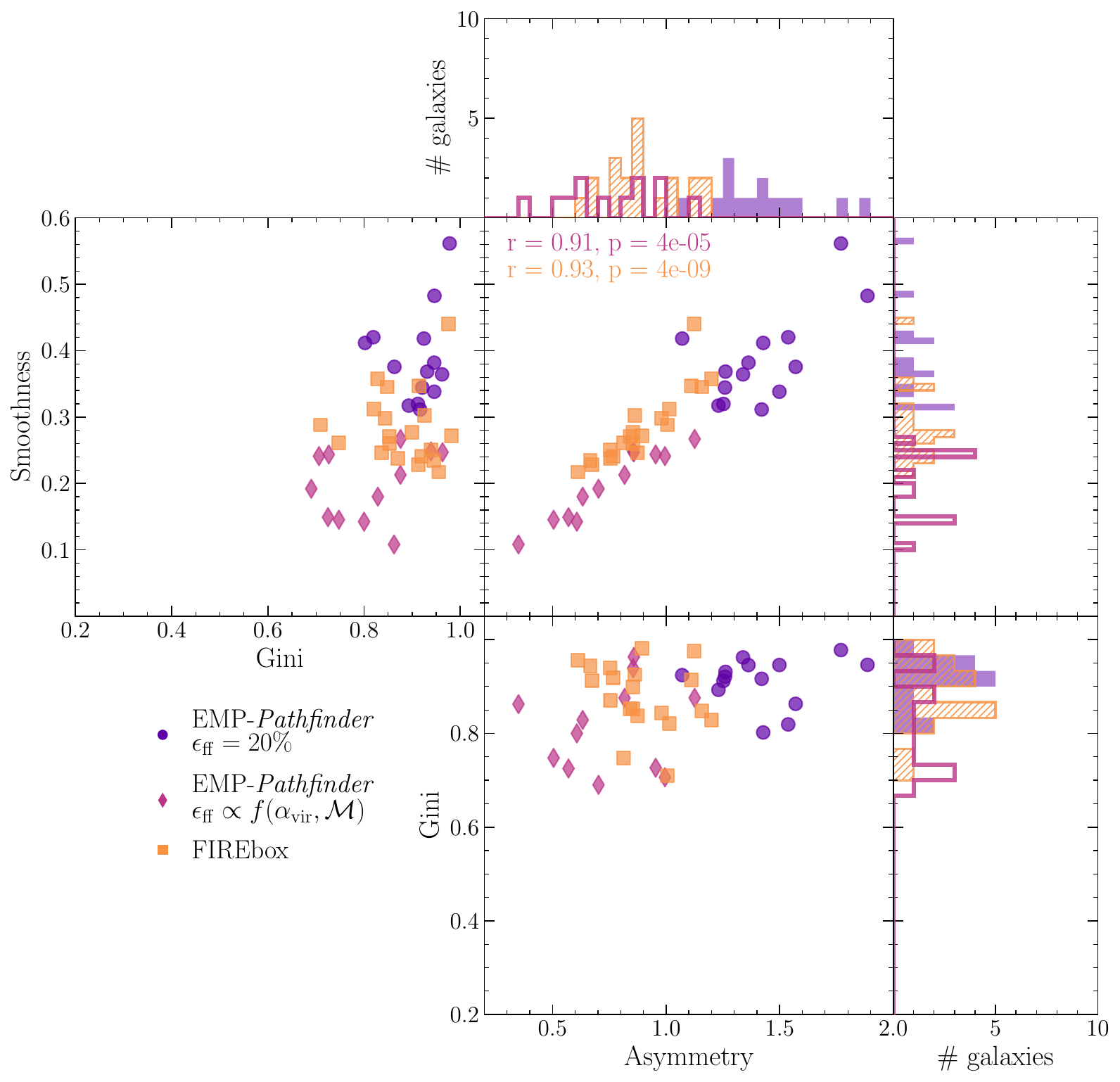}
    \caption{\hi morphology as classified by the non-parametric morphological indicators Smoothness, Gini and Asymmetry (plotted against each other in the three main panels), measured from face-on \hi surface density projections of the galaxies with a resolution of 80 pc and column density threshold of $3\times10^{20}~\rm{cm}^{-2}$. Purple circles denote the \empc~galaxies, while magenta diamonds denote the \empv~galaxies, and orange squares denote the \fc box galaxies. Histograms show the marginal distributions for each indicator and galaxy sub-set. If a statistically significant correlation between two non-parametric morphology indicators is present, the Spearman rank correlation coefficient and $p$-value are included in the panels, colour indicating which simulated galaxy sub-set they apply to. The galaxies simulated with different sub-grid physics occupy distinctly different parts of the Asymmetry-Smoothness parameter space.}
    \label{fig:HImorph_hc}
\end{figure}

In this Appendix, we show the non-parametric morphological parameter measurements computed from maps with a higher density column threshold of $3\times10^{20}~\cm^{-2}$, coarser resolution and random inclinations in Figures~\ref{fig:HImorph_hc} {--}~\ref{fig:HImorph_inc70}.
Comparing Figure~\ref{fig:HImorph} with Figure~\ref{fig:HImorph_hc}, shows that a higher column density cut does not affect the trends of the Asymmetry measurements strongly. It does lead to larger rotational asymmetries, the differences are more pronounced for the \fc box and \empc~galaxies (the smooth and symmetric sub-set of \empv~galaxies remains very smooth and symmetric), but differences are of order $\sim$20~per~cent. The Smoothness parameter is far more affected by the density cut. The difference is strongest for the \empc~galaxies, with Smoothness increasing by approximately a factor of 2 across the sample, i.e.\ the resultant maps are far less smooth than the fiducial ones. \fc box galaxies are less affected, the Smoothness increases by a factor of 1.5-2 for 5 galaxies only. The remainder of the \fc box galaxies and the \empv~galaxies experiences an order 0.01 increase in Smoothness. Gini remains largely unaffected by the increased surface density threshold. The scatter is slightly reduced, with the minimum values increasing from $\sim$0.60 to 0.65. 

\begin{figure}
    \centering
    \includegraphics[width=\linewidth]{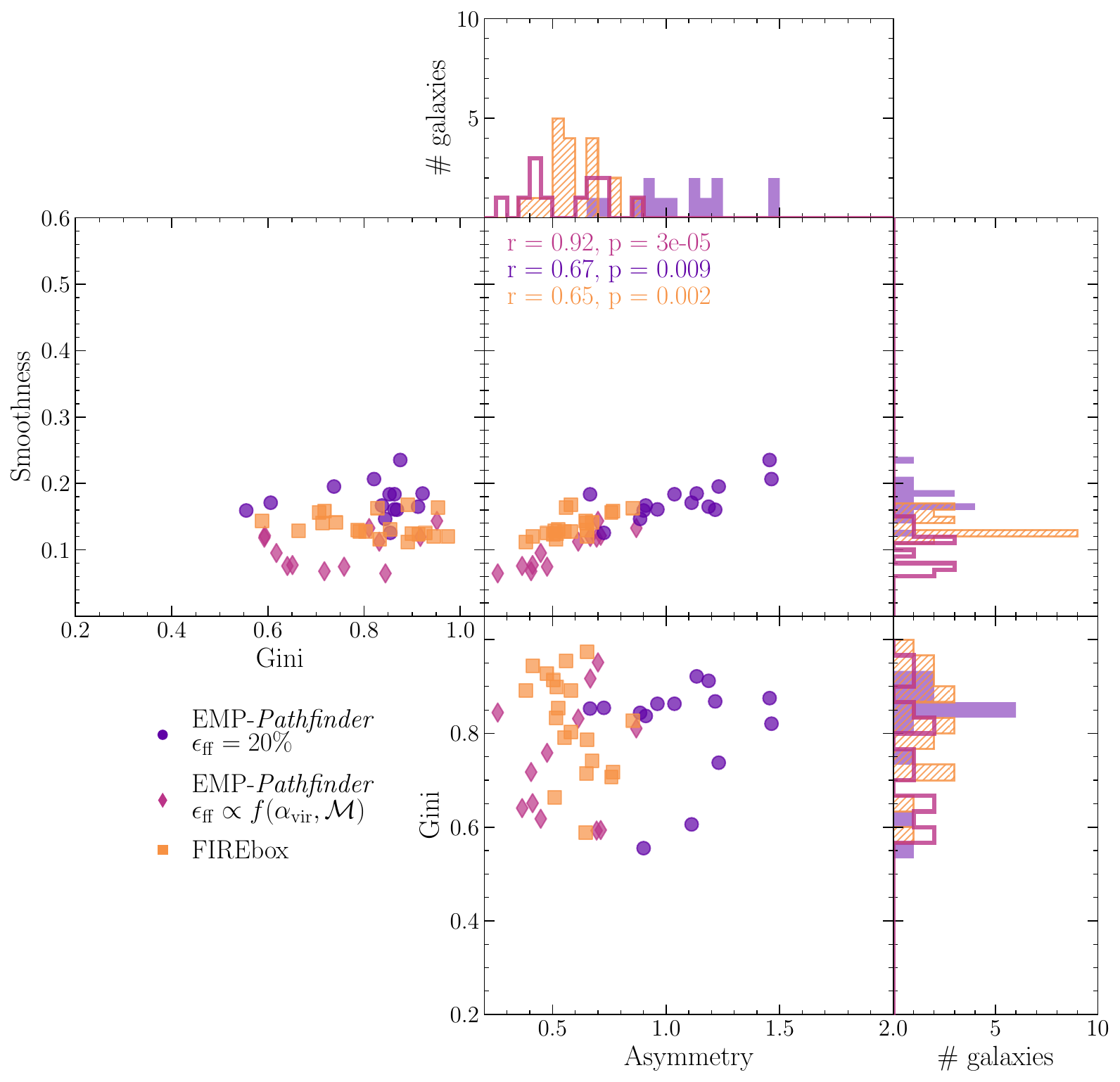}
    \caption{\hi morphology as classified by the non-parametric morphological indicators Smoothness, Gini and Asymmetry (plotted against each other in the three main panels), measured from face-on \hi surface density projections of the galaxies with a resolution of 500 pc and column density threshold of $7\times10^{19}~\rm{cm}^{-2}$. Purple circles denote the \empc~galaxies, while magenta diamonds denote the \empv~galaxies, and orange squares denote the \fc box galaxies. Histograms show the marginal distributions for each indicator and galaxy sub-set. If a statistically significant correlation between two non-parametric morphology indicators is present, the Spearman rank correlation coefficient and $p$-value are included in the panels, colour indicating which simulated galaxy sub-set they apply to. The galaxies simulated with different sub-grid physics occupy distinctly different parts of the Asymmetry-Smoothness parameter space.}
    \label{fig:HImorph_500pc}
\end{figure}

\begin{figure}
    \centering
    \includegraphics[width=\linewidth]{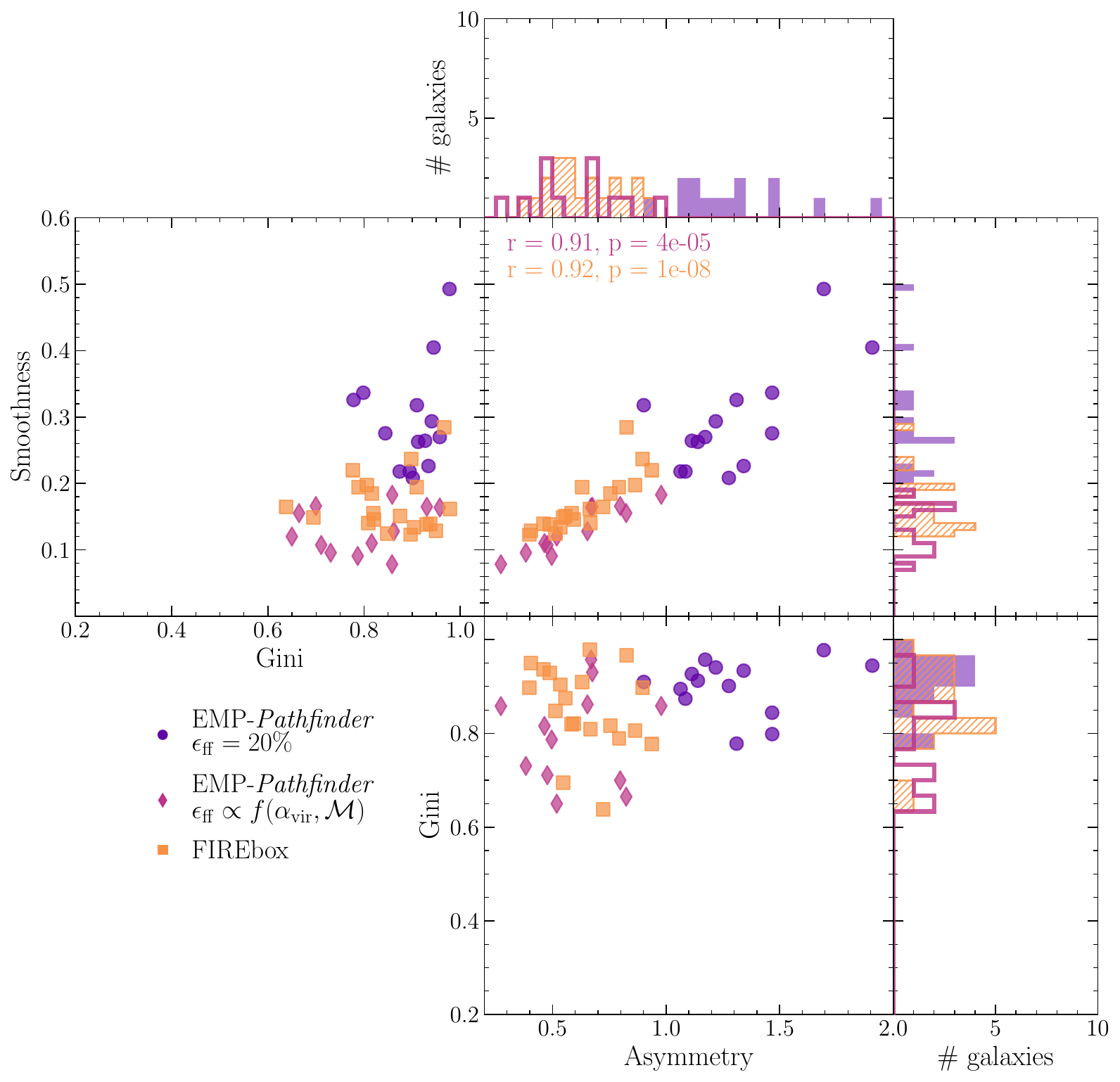}
    \caption{\hi morphology as classified by the non-parametric morphological indicators Smoothness, Gini and Asymmetry (plotted against each other in the three main panels), measured from face-on \hi surface density projections of the galaxies with a resolution of 500 pc and column density threshold of $3\times10^{20}~\rm{cm}^{-2}$. Purple circles denote the \empc~galaxies, while magenta diamonds denote the \empv~galaxies, and orange squares denote the \fc box galaxies. Histograms show the marginal distributions for each indicator and galaxy sub-set. If a statistically significant correlation between two non-parametric morphology indicators is present, the Spearman rank correlation coefficient and $p$-value are included in the panels, colour indicating which simulated galaxy sub-set they apply to. The galaxies simulated with different sub-grid physics occupy distinctly different parts of the Asymmetry-Smoothness parameter space.}
    \label{fig:HImorph_500pc_hc}
\end{figure}

To test the effect of resolution, we smoothed the maps to a resolution of 500 pc using a Gaussian smoothing kernel, and subsequently regridded them such that the full-width-half-maximum still corresponds to approximately 4 pixels prior to computing the non-parametric morphological indicators. Figures~\ref{fig:HImorph_500pc} and~\ref{fig:HImorph_500pc_hc} show the results for the fiducial and the higher column density cut, respectively. The coarser resolution makes the gas distributions smoother and more symmetric, i.e.\ results in $\sim$50~per~cent smaller Smoothness, and lower Asymmetry values. The change in Asymmetry is most pronounced for \fc box galaxies, where the Asymmetry decreases by up to 30~per~cent. Since the Smoothness and Asymmetry \empv~galaxies are the least affected by the smoothing, it increases the overlap between \fc box and \empv~galaxies. 

\begin{figure}
    \centering
    \includegraphics[width=\linewidth]{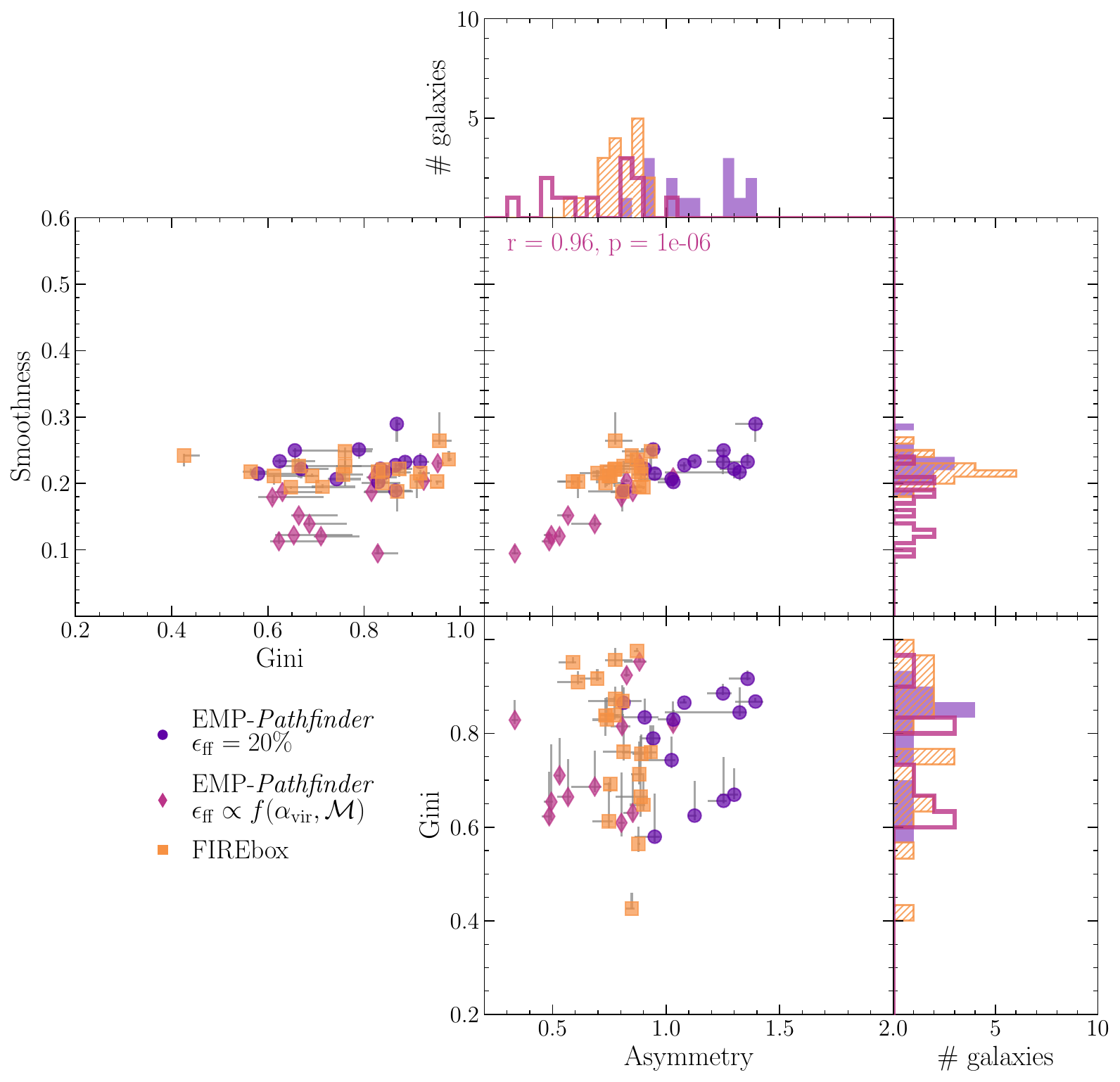}
    \caption{\hi morphology as classified by the median non-parametric morphological indicators Smoothness, Gini and Asymmetry (plotted against each other in the three main panels), measured from face-on \hi surface density projections of the galaxies with a resolution of 80 pc and column density threshold of $7\times10^{19}~\rm{cm}^{-2}$ and inclinations $<70^{\circ}$. Purple circles denote the \empc~galaxies, while magenta diamonds denote the \empv~galaxies, and orange squares denote the \fc box galaxies. Histograms show the marginal distributions for each indicator and galaxy sub-set. Error bars range from the minimum to the maximum of each non-parametric morphological indicator, for the inclinations considered.  If a statistically significant correlation between two non-parametric morphology indicators is present, the Spearman rank correlation coefficient and $p$-value are included in the panels, colour indicating which simulated galaxy sub-set they apply to. The galaxies simulated with different sub-grid physics occupy distinctly different parts of the Asymmetry-Smoothness parameter space.}
    \label{fig:HImorph_inc}
\end{figure}

\begin{figure}
    \centering
    \includegraphics[width=\linewidth]{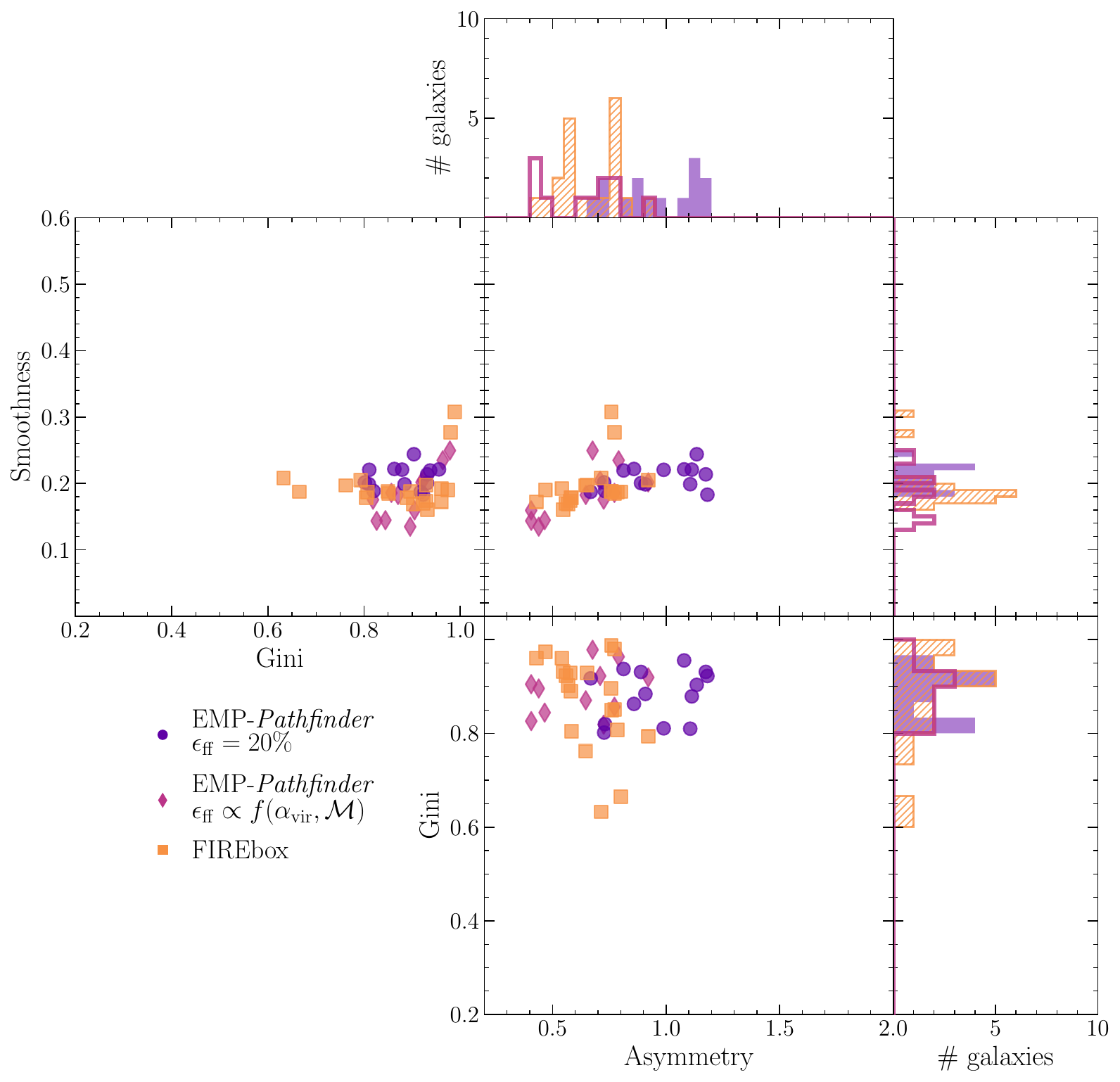}
    \caption{\hi morphology as classified by the median non-parametric morphological indicators Smoothness, Gini and Asymmetry (plotted against each other in the three main panels), measured from \hi surface density projections of the galaxies with a resolution of 80 pc and column density threshold of $7\times10^{19}~\rm{cm}^{-2}$ and inclination $i=70^{\circ}$. Purple circles denote the \empc~galaxies, while magenta diamonds denote the \empv~galaxies, and orange squares denote the \fc box galaxies. Histograms show the marginal distributions for each indicator and galaxy sub-set. No statistically significant correlations between two non-parametric morphology indicators are present. Galaxies appear more symmetric at high inclinations and the overlap between the different physics sub-sets increases significantly.}
    \label{fig:HImorph_inc70}
\end{figure}

Finally, we test the effect of inclination. Figure~\ref{fig:HImorph_inc} shows the median Gini, Asymmetry and Smoothness measurements for each simulated galaxy computed on the fiducial maps with inclinations $< 70^{\circ}$. Grey error bars range from the minimum to the maximum value. Gini is the most affected by changes in inclination, whilst Smoothness and Asymmetry only vary a little. This no longer holds true for $i\geq70^{\circ}$, as is shown in Figure~\ref{fig:HImorph_inc70} for our fiducial density threshold and resolution. At these large inclinations, the galaxies appear more symmetric and there are no longer any statistically significant correlations between the non-parametric morphological indcators. Furtheremore, there is more overlap between the galaxy sub-sets simulated with different physics in Asymmetry and Smoothness, compared to lower inclinations.  

\bsp	
\label{lastpage}
\end{document}